\documentclass[prd,preprint,superscriptaddress,amsmath,amssymb,nofootinbib]{revtex4-1}
\usepackage{graphicx}
\usepackage{dcolumn}
\usepackage{bm}
\usepackage{amssymb}
\usepackage{amsmath}
\usepackage{epsfig}    
\usepackage{color}
\usepackage{slashed}
\usepackage{hhline}
\usepackage{xcolor}
\usepackage{enumitem}
\usepackage[colorlinks=true,linkcolor=blue,citecolor=blue,urlcolor=blue]{hyperref}
\usepackage{xcolor}
\usepackage{cancel}
\usepackage[normalem]{ulem}
\def\be{\begin{equation}}
\def\ee{\end{equation}}
\newcommand{\bea}{\begin{eqnarray}}
\newcommand{\eea}{\end{eqnarray}}

\begin{document}


\title{Signatures of Long-Lived Heavy Neutral Leptons from Neutrinophilic Charged Higgs Pair Production at the LHC}

\author{Nobuchika Okada}
\email{okadan@ua.edu}
\affiliation{Department of Physics and Astronomy,
University of Alabama, Tuscaloosa, USA}

\author{Prasenjit Sanyal}
\email{prasenjit.sanyal01@gmail.com}
\affiliation{Center for Quantum Spacetime, Sogang University, 35 Baekbeom-ro, Seoul 121-742, South Korea}
\affiliation{Department of Physics, Sogang University, Seoul 121-742, South Korea}

\author{Ravindra Kumar Verma}
\email{ravindra.kumar.verma@cern.ch}
\affiliation{Helsinki Institute of Physics, P.O. Box 64, 00014 University of Helsinki, Finland}

\date{\today}

\begin{abstract}
	
 In the neutrinophilic Higgs doublet framework, the neutrino Dirac Yukawa couplings can be sizable because of the small vacuum expection value of the extra Higgs doublet, even for a low seesaw scale. Due to this structure, the neutrinophilic charged Higgs bosons, once created, decay dominantly into heavy neutral leptons (HNLs) and charged leptons. This is a new mechanism to produce a gauge singlet HNL without suppressed cross sections. In the standard seesaw, one HNL can be long-lived, when the lightest neutrino is sufficiently light. We investigate displaced vertex signatures of the long-lived HNLs produced from the decays of the charged Higgs pair at the high luminosity LHC. We consider one displaced vertex as well as two displaced vertices signatures and perform a dedicated simulation to identify the displaced leptons. \textcolor{black}{We find that high statistical significance can be achieved for the observation of one displaced vertex for charged Higgs pair production cross section $>\mathcal{O}(1)$ fb. On the other hand, the observation of two displaced vertices is challenging, particularly for charged Higgs pair production cross section below $\mathcal{O}(10)$ fb.}

\end{abstract}

\preprint{CQUeST-2026-0777}

\maketitle

\section{Introduction}
\label{Introduction}

The observation of neutrino oscillations~\cite{ParticleDataGroup:2024cfk} has firmly established that neutrinos are massive 
and that lepton flavors mix, providing clear evidence for physics beyond the Standard Model (SM). 
Despite its remarkable success, the SM cannot accommodate non-zero neutrino masses, and 
understanding their origin remains one of the most important open problems in particle physics.

Among various proposals, the type-I seesaw mechanism~\cite{Minkowski:1977sc, Yanagida:1979as, GellMann:1980vs, Mohapatra:1979ia}
offers one of the simplest and most well-motivated frameworks to generate tiny neutrino masses. 
In this framework, heavy right-handed neutrinos (RHNs), which are singlets under the SM gauge group, are introduced 
along with their Majorana mass terms. 
After electroweak symmetry breaking, the interplay between Dirac and Majorana mass terms naturally 
yields suppressed light neutrino masses. 
A direct confirmation of this mechanism requires the production and detection of RHNs 
(which we refer to as heavy neutral leptons (HNLs), corresponding to the mass eigenstates after the seesaw mechanism) at collider experiments. 
In particular, collider searches can probe their Majorana nature through lepton number violating signatures 
(see, for example, Ref.~\cite{Atre:2009rg}), 
as well as test their flavor structure against neutrino oscillation data.

Extensive studies have been devoted to HNL production at high-energy colliders. 
However, since RHNs are singlets under the SM gauge group, HNL production is typically suppressed from a theoretical perspective. 
One possibility is through electroweak interactions induced by the mixing between light and heavy neutrinos, 
which generally leads to small production cross sections. 
Even when the production rate is small, a long-lived HNL can still be detectable over the SM background 
through displaced vertex signatures. 
Prospects for discovering long-lived HNLs at the Large Hadron Collider (LHC) have been extensively investigated 
for HNL masses below the $W$ boson mass~\cite{Gago:2015vma,Accomando:2016rpc,Abada:2018sfh,Cottin:2018nms,Cottin:2018kmq,Drewes:2019fou,Liu:2019ayx,  
Beltran:2021hpq,Liu:2022ugx,Yang:2024nmk}, 
as well as for heavier HNLs~\cite{Helo:2013esa,Accomando:2017qcs,Chiang:2019ajm,Bandyopadhyay:2022mej,Lavignac:2025fjt}. 
Similar studies for future colliders, such as the 100 TeV hadron collider FCC-hh and future lepton colliders, 
can be found in Refs.~\cite{Padhan:2022fak,Liu:2023klu} and Refs.~\cite{Antusch:2016vyf,Blondel:2022qqo,Urquia-Calderon:2023dkf,Bi:2024pkk,Aleksan:2024hyq}, respectively. 
Experimental searches for long-lived HNLs at the LHC have been reported by the ATLAS~\cite{ATLAS:2019kpx,ATLAS:2022atq} 
and CMS~\cite{CMS:2022fut,CMS:2023jqi,CMS:2024hik} collaborations.

Another possibility to produce HNLs at high-energy colliders arises in extensions of the SM with additional mediators. 
In certain classes of new physics models, such as gauged $B-L$ (baryon number minus lepton number) or more general $U(1)_X$ extensions, 
new particles such as gauge bosons or scalar fields can mediate interactions between HNLs and SM fermions. 
Through these mediators, HNLs can be produced with sizable cross sections at colliders~\cite{Kang:2015uoc,Cox:2017eme,Accomando:2017qcs,Das:2017flq,Das:2017deo,Jana:2018rdf,Das:2018tbd}. 

In this paper, we explore an alternative production mechanism for HNLs 
within the framework of the so-called neutrinophilic Higgs doublet model~\cite{Ma:2000cc,Wang:2006jy,Gabriel:2006ns,Davidson:2009ha,Haba:2010zi,Haba:2011nb,Haba:2012ai}. 
In this model, an additional Higgs doublet is introduced that couples exclusively to the SM lepton doublets and RHNs. 
This structure allows the vacuum expectation value (VEV) of the extra Higgs doublet to be much smaller than that of the SM Higgs doublet, 
thereby generating naturally small Dirac neutrino masses without requiring extremely small Yukawa couplings, 
even for relatively low seesaw scales. 
  \textcolor{black}{Furthermore, the VEV hierarchy leads to a very small mixing between the two Higgs doublets and  due to the neutrinophilic nature of the extra Higgs doublet, the additional Higgs bosons evade stringent constraints from hadron collider experiments and flavor physics.}
The direct bounds are therefore relatively weak, with the LEP experiment setting a lower limit of approximately 
$m_{H^\pm} \gtrsim 80~\mathrm{GeV}$~\cite{ALEPH:2013htx}. \textcolor{black}{The Higgs boson originating primarily from the SM Higgs doublet can be identified as the observed 125 GeV Higgs boson, with couplings similar to that of the SM Higgs boson.}

Within this model setup, we focus on a novel production channel for HNLs via the decay of neutrinophilic charged Higgs bosons. 
At the LHC, charged Higgs bosons can be pair-produced through the Drell--Yan process.  
In the limit of small mixing between the two Higgs doublets, the extra Higgs bosons \textcolor{black}{can nearly be} degenerate in mass, 
which suppresses the bosonic decay modes such as \textcolor{black}{$H^\pm \to W^\pm H/W^\pm A$ and $H\to ZA$}, 
where $H$ and $A$ denote the CP-even and CP-odd neutral Higgs bosons, respectively.  
\textcolor{black}{Similarly, the decay mode into the SM Higgs boson, $H^\pm \to W^\pm h$ and $A\to Zh$, are also suppressed. In addition, the mass degeneracy among the extra Higgs bosons suppresses the contributions to the electroweak oblique parameters $S$, $T$, and $U$, thereby satisfying the  electroweak precision constraints~\cite{Grimus:2007if, Grimus:2008nb, Bahl:2021str}.}  
As a result, the dominant decay channel of the charged Higgs boson is into a charged lepton and a HNL, 
provided it is kinematically allowed. 
\textcolor{black}{In this work, we consider the parameter region where $m_{H^\pm}, m_{N} > m_W$ and 
so that the HNLs produced from charged Higgs decays can subsequently undergo on-shell decays into SM gauge bosons.}

A particularly interesting aspect of this setup is the possibility of realizing long-lived HNLs. 
Most existing studies focus on relatively light HNLs with masses below the $W$ boson mass, 
for which the decays proceed via off-shell gauge bosons, leading to long lifetimes. 
However, as shown in Ref.~\cite{Das:2019fee}, within the type-I seesaw framework with three RHNs, 
one of the HNLs can become long-lived even when it is heavy and decays via on-shell gauge bosons. 
In this case, the lifetime is inversely proportional to the lightest neutrino mass while remaining consistent with neutrino oscillation data. 
In this paper, we investigate the production of such long-lived HNLs via charged Higgs pair production 
and analyze their displaced vertex signatures at the LHC. 
We demonstrate that this mechanism provides a viable and complementary avenue to probe the origin of neutrino masses 
and the structure of the seesaw mechanism. 

The paper is organized as follows. In Sec.~\ref{Sec:Simple Setup}, we present an overview of the neutrinophilic Higgs doublet model and discuss how a long-lived HNL state can arise in both NH and IH scenarios within the standard seesaw mechanism, in contrast to the minimal seesaw. In Sec.~\ref{Sec: Phenomenology}, we obtain bounds on the vacuum expectation value of the neutrinophilic Higgs doublet from the lepton flavor violating $\mu \to e\gamma$ process and then we discuss the branching ratios of the neutrinophilic charged Higgs to the long-lived HNL states, followed by the branching ratios of the HNLs in the NH and IH scenarios. In the same section, we show the proper decay lengths of the long-lived HNLs in NH and IH, as a function of the lightest neutrino mass. In Sec.~\ref{Sec: Collider Simulation and Analysis}, we discuss the displaced vertex signatures of the long-lived HNLs arising from the decays of the neutrinophilic charged Higgs at the high luminosity LHC (HL-LHC), by taking into account the details of the collider simulation. Our results are presented in Sec.~\ref{Sec: Results}. Finally, we conclude in Sec.~\ref{Sec: Conclusions}.

\section{Model Setup}
\label{Sec:Simple Setup}

\textcolor{black}{{The canonical seesaw or the type-I seesaw is the most widely accepted formalism for neutrino mass generation. Under this mechanism, the neutrino mass matrix can be written as 
\begin{eqnarray}
\mathcal{M} = \left(\begin{array}{cc}
0 & m_D \\
m_D^T & m_N\\
\end{array}\right),
\label{Eq: Neutrino mass term}
\end{eqnarray} 
where $m_D$ is the Dirac mass matrix, and $m_N$ is the Majorana mass matrix. Without any loss of generality we can consider $m_N$ to be diagonal and the small neutrino mass requires the seesaw hierarchy $m_N \gg m_D$. This small Dirac mass term can be achieved from a small vacuum expectation value (VEV) of the neutrinophilic Higgs doublet $H_2$. Following Ref.~\cite{Ma:2000cc}, we consider a simple model setup, where a global lepton number $U(1)_L$ symmetry is imposed, under which $H_1:L=0$, $H_2:L=-1$ and the SM singlet RHNs $N_R: L = 0$.  Under these charge assignments we can have an additional term in the Yukawa Lagrangian    
\begin{eqnarray}
-\mathcal{L}_y \supset  y_D^{ij} \bar{L}_{Li} \widetilde{H}_2 N_{Rj} + h.c.,
\label{Eq: Yukawa Lagrangian}
\end{eqnarray} 
where $\widetilde{H}_2 = i\sigma_2 H^*_2$
and the charged leptons and quarks receive their masses from the VEV of the SM Higgs doublet $H_1$.
The scalar potential can be written as 
 \begin{eqnarray}
 	V &=& m_{11}^2 H_1^\dagger H_1 + m_{22}^2 H_2^\dagger H_2 - m_{12}^2\Big[ H_1^\dagger H_2 + h.c. \Big] + \frac{\lambda_1}{2} (H_1^\dagger H_1)^2 +  \frac{\lambda_2}{2} (H_2^\dagger H_2)^2  \nonumber \\
 	&+& \lambda_3 (H_1^\dagger H_1) (H_2^\dagger H_2) + \lambda_4 (H_1^\dagger H_2)(H_2^\dagger H_1),
 	\label{Eq: Scalar Potential}
 \end{eqnarray}
 where the lepton number symmetry is softly broken by $m_{12}^2 \neq 0$ and the Higgs doublets can be parameterized as
 \begin{eqnarray}
 	H_i = \left(\begin{array}{cc}
 		\varphi_i^+  \\
 		\frac{v_i + \rho_i + i \eta_i}{\sqrt{2}}\\
 	\end{array}\right),
 	\label{Eq: Higgs Doublets}
 \end{eqnarray} 
 with $v_1$ and $v_2$ as the VEVs of the Higgs doublets $H_1$ and $H_2$, and the electroweak (EW) VEV is given by $v=\sqrt{v_1^2 + v_2^2} \simeq 246$ GeV.  The VEV hierarchy $v_2 \ll v_1$ implies that the SM gauge symmetry is broken spontaneously at the electroweak (EW) scale by the VEV of $H_1$, i.e. $v_1 \simeq v$.
}}

 \textcolor{black}{{The quartic couplings of the scalar potential should be
 $|\lambda_i| < 4\pi$ to satisfy the perturbativity condition. The vacuum stability requires the potential to be bounded from below, and this gives us the constraints \cite{Gunion:2002zf}
  \begin{eqnarray}
  	\lambda_1, \lambda_2 > 0, \quad \lambda_3 + \sqrt{\lambda_1 \lambda_2} > 0, \quad \lambda_3 + \lambda_4 +\sqrt{\lambda_1 \lambda_2} > 0.
  \end{eqnarray}
  The VEVs of the Higgs doublets are obtained my minimizing the scalar potential in Eq.~[\ref{Eq: Scalar Potential}] 
  \begin{eqnarray}
  	v_1\Big [ m_{11}^2 + \frac{1}{2} v_1^2 \lambda_1 + \frac{1}{2} v_2^2 (\lambda_3 + \lambda_4) \Big] - m_{12}^2 v_2 &=& 0, \nonumber \\
  	v_2 \Big[ m_{22}^2 + \frac{1}{2} v_2^2 \lambda_2 + \frac{1}{2} v_1^2 (\lambda_3 + \lambda_4) \Big] - m_{12}^2 v_1 &=& 0.
  	\label{Eq: minimization of potential}
  \end{eqnarray}
  In the limit $v_2\ll v_1$, and assuming $m_{11}^2<0$, $m_{22}^2>0$, and $|m_{12}^2|\ll m_{22}^2$, the minimization conditions lead to
  \begin{eqnarray}
  	v_1 \simeq \sqrt{-\frac{2m_{11}^2}{\lambda_1}}, \quad v_2 \simeq \frac{m_{12}^2 v_1}{m^2_{22} + \frac{1}{2} v_1^2 (\lambda_3 + \lambda_4)}.
  \end{eqnarray}
 For $v_2 $ of the order of few MeV and $m_{22}$ ranging from EW scale to 1 TeV, we obtain $|m_{12}^2| <$ 100 GeV$^2$.  The smallness of $|m_{12}^2|$ is more natural and implies the soft explicit breaking of the lepton number symmetry. 
}}
 
 \textcolor{black}{{The VEV hierarchy also induces very small mixing between the two Higgs doublets. As a result, the SM Higgs mass eigenstate comes predominantly from the SM Higgs doublet $H_1$, while the non-SM Higgs bosons originate mainly from the neutrinophilic Higgs doublet $H_2$. The physical scalar spectrum consists of a charged Higgs pair $H^\pm$, a pseudoscalar $A$, and two CP-even Higgs bosons, $h$ and $H$, where $h$ is identified with the observed  Higgs boson.  The Higgs boson mass eigenstates are given as 
 \begin{eqnarray}
 	&&H^\pm = \frac{v_1 \varphi^\pm_2 - v_2 \varphi^\pm_1}{\sqrt{v_1^2 + v_2^2}}, \quad A = \frac{v_1 \eta_2 - v_2 \eta_1}{\sqrt{v_1^2 + v_2^2}}, \nonumber \\
 	&&h \simeq \frac{v_1 \rho_1 + v_2 \rho_2}{\sqrt{v_1^2 + v_2^2}} ,  \quad H \simeq \frac{v_1 \rho_2 - v_2 \rho_1}{\sqrt{v_1^2 + v_2^2}},
\end{eqnarray} 
 and the masses are
 \begin{eqnarray}
 	m_{H^\pm}^2 &\simeq& m_{22}^2 + \frac{1}{2} v_1^2 \lambda_3, \nonumber \\
 	m_{A}^2 &\simeq&  m_{22}^2 + \frac{1}{2} v_1^2 (\lambda_3 + \lambda_4), \nonumber \\
 	m_{h}^2 &\simeq& v_1^2 \lambda_1, \nonumber \\
 	m_{H}^2 &\simeq&  m_{22}^2 + \frac{1}{2} v_1^2 (\lambda_3 + \lambda_4), 
 	\label{Eq: Higgs Masses}
 \end{eqnarray}	
 where we neglect terms $\mathcal{O}(v^2_2)$ and $\mathcal{O}(m_{12}^2 v_2/v_1)$. The Higgs boson $h$ has properties similar to those of the SM Higgs boson, and its couplings to the charged fermions and gauge bosons are SM-like. The additional  Higgs bosons  $H^\pm$, $A$ and $H$ are neutrinophilic. As shown in Eq.~(\ref{Eq: Higgs Masses}), $m_H \simeq m_A$. Furthermore, in the limit $\lambda_4 \simeq 0$, the non-SM neutrinophilic Higgs bosons become nearly degenerate in mass.
}}
 
\textcolor{black}{{In this model setup, the RHNs are simply added with Majorana masses $m_{N}$.  The Dirac mass term can be written as }}
\begin{eqnarray}
m_D = \frac{y_D}{\sqrt{2}}v_2 ,
\end{eqnarray}
and the mass term from the Yukawa Lagrangian is
\begin{eqnarray}
-\mathcal{L}_Y^{\text{mass}}= \frac{1}{2}\left(
\begin{array}{c}
\bar{\nu}_L \\ \bar{N}_R^C 
\end{array}\right)^T
\mathcal{M}\left(
\begin{array}{c}
\nu^C_L \\ N_R 
\end{array}\right)
 + h.c. ,
\end{eqnarray} 
where $\mathcal{M}$ is given in Eq.[\ref{Eq: Neutrino mass term}]. Diagonalizing the mass matrix gives the seesaw formula, which can be written approximately as 
\begin{eqnarray}
m_\nu &=& -m_D m_N^{-1} m_D^T, \nonumber \\
R &=& m_D m_N^{-1}, 
\label{Eq: Seesaw formula} 
\end{eqnarray} 
where $R$ is the light-heavy neutrino mixing matrix, $m_\nu$ is the light neutrino matrix and $m_N =$ diag $({m_{N_1}, m_{N_2}, m_{N_3}})$ are the masses of the heavy neutral leptons (HNL). The $m_\nu$ mass matrix can be diagonalized by the unitary PMNS matrix $U$ such that 
\begin{eqnarray}
D_\nu = U^T m_{\nu} U = \text{diag}(m_{1}, m_{2}, m_{3}).
\end{eqnarray}
We can express the light neutrino flavor eigenstates $\nu_\alpha$ to the mass eigenstates of light neutrinos $\nu_i$ and HNLs $N_i$ by  
\begin{eqnarray}
\nu_\alpha = N_{\alpha i}\nu_i + R_{\alpha i} N_i ,
\end{eqnarray}
where $N = \Big(1 - \frac{1}{2} \epsilon \Big)U$ and $\epsilon =  R^* R^T$ is the small non-unitarity parameter. The Dirac mass matrix $m_D$ can be parametrized in terms of $D_\nu$ and $m_N$ as \cite{Casas:2001sr}
\begin{eqnarray}
m_D = U^* \sqrt{D_\nu} \mathcal{O} \sqrt{m_N} ,
\label{Eq: Cassas Ibarra}
\end{eqnarray}  
where $\mathcal{O}$ is an orthogonal matrix, and the PMNS matrix is given by
\begin{eqnarray}
U = \left(\begin{array}{ccc}
c_{12}c_{13} & s_{12}c_{13} & s_{13}e^{i\delta} \\
-s_{12}c_{23} - c_{12}s_{23}s_{13}e^{i\delta} & c_{12}c_{23} - s_{12}s_{23}s_{13}e^{i\delta} & s_{23}c_{13} \\
s_{12}c_{23} - c_{12}c_{23}s_{13}e^{i\delta} & -c_{12}s_{23} - s_{12}c_{23}s_{13}e^{i\delta} & c_{23}c_{13}\\
\end{array}\right)
\left(\begin{array}{ccc}
1 & 0 & 0 \\
0 & e^{i\rho_1} & 0\\
0 & 0 & e^{i\rho_2} \\
\end{array}\right).
\end{eqnarray}
Here $c_{ij} = \cos{\theta_{ij}},~s_{ij}=\sin{\theta_{ij}}$, $\delta$ is the Dirac $CP$-phase and $\rho_{1,2}$ are the Majorana phases.
In the following analysis we adopt the best-fit values of the neutrino oscillation data for the parameters
 $\Delta m_{12}^2 = m_2^2 - m_1^2 = 7.53 \times 10^{-5}$ eV$^2$, $\Delta m_{23}^2 = |m_3^2 - m_2^2| = 2.45 \times 10^{-3}$ eV$^2$, $s^2_{\theta_{12}} = 0.307$, $s^2_{\theta_{23}} = 0.558$, $s^2_{\theta_{13}} = 2.19\times 10^{-2}$ and $\delta = 1.19\pi$.    
These measurements leave two possible mass hierarchies undetermined, the normal hierarchy (NH) where the light neutrino masses are in order $m_1 < m_2 < m_3$, and the inverted hierarchy (IH) where the light neutrino mass eigenstates are ordered as $m_3 < m_1 < m_2$. Therefore, the lightest neutrino mass becomes an independent variable. We require $\sum_i m_i < 0.12$ eV to be consistent with the Planck upper limit on the sum of neutrino masses.
In the NH, $m_{\text{lightest}} = m_1$ and $D_\nu^{\text{NH}}$ = diag ($m_{\text{lightest}}, m_2^{\text{NH}}, m_3^{\text{NH}}$) with $m_2^{\text{NH}} = \sqrt{\Delta m_{12}^2 + m^2_{\text{lightest}}}$ and $m_3^{\text{NH}} = \sqrt{\Delta^2_{23} + (m_2^{\text{NH}})^2}$. 
In the case of IH, $m_{\text{lightest}} = m_3$ and $D_\nu^{\text{IH}}$ = diag ($ m_1^{\text{IH}}, m_2^{\text{IH}}, m_{\text{lightest}}$) with $m_2^{\text{IH}} = \sqrt{\Delta m^2_{23} + m^2_{\text{lightest}}}$
and $m_1^{\text{IH}} = \sqrt{(m_2^{\text{IH}})^2 - \Delta m_{12}^2}$. In both cases, we can write the HNL states as $m_N =$ diag $({m_{N_1}, m_{N_2}, m_{N_3}})$.

Since $m_{\text{lightest}}$ is a free parameter, we consider a special situation with $m_{\text{lightest}} = 0$. This is the so-called minimal seesaw, in which only two RHNs participate in the seesaw mechanism. Before going into the details of how these two mechanisms differ in terms of producing long-lived HNLs at the colliders, we show the two-body partial decay widths of HNLs, which hold true independent of the neutrino mass hierarchy.

\textcolor{black}{The decay widths of HNLs are\footnote{\textcolor{black}{The small mixing ($\simeq v_2/v$) between the SM Higgs and the neutrinophilic Higgs gives rise to the decay process $N_i \to \nu_\alpha h$, if kinematically allowed.}}
\begin{eqnarray}
	\Gamma(N_i \to \ell_{\alpha} W) &=& \frac{|R_{\alpha i}|^2}{8 \pi} \frac{(m^2_{N_i} - m^2_{W})^2 (m^2_{N_i} + 2 m^2_{W})}{m_{N_i}^3 v^2}, \nonumber \\
	\Gamma (N_i \to \nu_\alpha Z) &=& \frac{|R_{\alpha i}|^2}{16 \pi} \frac{(m^2_{N_i} - m^2_{Z})^2 (m^2_{N_i} + 2 m^2_{Z})}{m_{N_i}^3 v^2}.
	\label{Eq: HNL decay widths}  \nonumber \\
	\Gamma (N_i \to \nu_\alpha h) &=& \frac{|R_{\alpha i}|^2}{16 \pi} \frac{(m_{N_i}^2 - m_h^2)^2}{m_{N_i}v^2}.
\end{eqnarray}   
Here, $\alpha$ is the SM lepton flavor index, $i$ is the HNL mass eigenstate index and $v \simeq 246$ GeV is the electroweak vacuum expectation value. The total decay width of $N_i$ is:
\begin{eqnarray}
	\Gamma_{N_i} = \sum\limits_{\alpha = e, \mu, \tau} \Gamma(N_i \to \ell_{\alpha} W) + \Gamma(N_i \to \nu_{\alpha} Z) +  \Gamma(N_i \to \nu_{\alpha} h),
\end{eqnarray}
}
and the proper decay length $c\tau_{N_i}$ can be written as 
\begin{eqnarray}
c\tau_{N_i}= \frac{1.97\times 10^{-13}}{\Gamma_{N_i}[\text{GeV}]} [\text{mm}],
\end{eqnarray}
where $\tau_{N_i}$ indicates the lifetime of the HNL, and $c$ is the speed of light.
To understand the long-lived nature of the HNLs, note that $\Gamma_{N_i}$ is proportional to 
\begin{eqnarray}
	\sum\limits_{\alpha = e, \mu, \tau} |R_{\alpha i}|^2 = (R^\dagger R)_{ii}. 
\end{eqnarray}

\subsection{Case I: Standard seesaw}
\label{Sec: Case I: Standard seesaw}

In the standard scenario, all three right handed neutrinos RHNs participate in the seesaw mechanism.  We have
\begin{eqnarray}
 D_{\nu} &=& \text{diag}(m_1, m_2,m_3), \nonumber \\
 \sqrt{D_\nu} &=&\text{diag}(\sqrt{m_1}, \sqrt{m_2},\sqrt{m_3}), \nonumber \\
 m_N  &=&\text{diag}(m_{N_1}, m_{N_2}, m_{N_3}), \nonumber \\
 m_N^{-1} &=& \text{diag}(1/m_{N_1}, 1/m_{N_2}, 1/m_{N_3}).
\end{eqnarray}	
  In Eq. [\ref{Eq: Cassas Ibarra}] the orthogonal $3\times 3$ matrix $\mathcal{O}$  can be written as
  \begin{eqnarray}
  	\mathcal{O} =  \left(\begin{array}{ccc}
  		1 & 0 & 0 \\
  		0 & c_x & s_x\\
  		0 & -s_x & c_x
  	\end{array}\right)
  	 \left(\begin{array}{ccc}
  		c_y & 0 & s_y \\
  		0 & 1 & 0\\
  		-s_y & 0 & c_y
  	\end{array}\right)
  	\left(\begin{array}{ccc}
  		c_z  & s_z & 0 \\
  		-s_z & c_z & 0\\
  		0 & 0 & 1
  	\end{array}\right),
  	\label{EQ: Orthogonal matrix 3 X 3}
  	\end{eqnarray}  
  	where the parameters $x,~y,~z$ are complex, in general.
  	
Considering the $R$ matrix given in Eq.~[\ref{Eq: Seesaw formula}] 
\begin{eqnarray}
    \label{Eq: R matrix}
	R &=& m_D m_N^{-1} = U^* \sqrt{D_\nu} \mathcal{O}\sqrt{m_N^{-1}}, \nonumber \\
	R^\dagger R &=& \sqrt{m^{-1}_N} \mathcal{O}^\dagger D_\nu \mathcal{O} \sqrt{m^{-1}_N}.
\end{eqnarray}  	 
For $\mathcal{O} = \mathbb{I}$ as an identity matrix, 
\begin{eqnarray}
    \label{Eq: RdR}
	 R^\dagger R &=& \sqrt{m^{-1}_N}  D_\nu  \sqrt{m^{-1}_N} = \text{diag}(m_1/m_{N1}, m_2/m_{N2}, m_3/m_{N3}) 
\end{eqnarray}
 In the NH, $m_{\text{lightest}} = m_1$ whereas in the IH, $m_{\text{lightest}} = m_3$. This formula indicates that $N_1,$ and $N_3$ can be long-lived in NH and IH scenarios, respectively, if $m_{\text{lightest}}$ is sufficiently small.

Exploring more in details about the $\mathcal{O} \neq \mathbb{I}$ case. If $x = y = 0$ and $z= z_1 + i z_2 \neq 0$ 
\begin{eqnarray}
	(R^\dagger R)_{11} &=& \frac{1}{2 m_{N_1}}\Big[ (m_1 - m_2) \cos 2z_1 + (m_1 + m_2)\cosh 2z_2 \Big] \nonumber \\
	&\simeq& \frac{1}{4 m_{N_1}}(m_1 + m_2) e^{2|z_2|} \quad\text{for}~ |z_2| \gg 1.  
\end{eqnarray}
Thus the decay width of $N_1$ can be exponentially enhanced for $\mathcal{O} \neq \mathbb{I}$. In the limit $z_1 = 0,~ z_2 = 0$, we get $(R^\dagger R)_{11} = m_1/m_{N_1}$ and therefore restores the long-lived nature of $N_1$.
For the case of $N_2$ decay width,
\begin{eqnarray}
	(R^\dagger R)_{22} &=& \frac{1}{2 m_{N_2}} \Big[ (m_2 - m_1)\cos 2 z_1 + (m_1 + m_2)\cosh 2 z_2 \Big] \nonumber \\
     &\simeq& \frac{1}{4 m_{N_2}}(m_1 + m_2)e^{2|z_2|}	\quad\text{for}~ |z_2| \gg 1,
\end{eqnarray}   	
and hence exponentially large decay width for $N_2$ when $\mathcal{O} \neq \mathbb{I}$. Once again in the limit $z_1 = 0,~ z_2 = 0$, we get $(R^\dagger R)_{22} = m_2/m_{N_2}$.
 For $N_3$ decay width, $(R^\dagger R)_{33} = m_3/m_{N_3}~ \forall~ z.$
 
 Now in the situation, if $x = z = 0$ and $y= y_1 + i y_2 \neq 0$.
 \begin{eqnarray}
 	(R^\dagger R)_{11} &=& \frac{1}{2 m_{N_1}}\Big[ (m_1 - m_3) \cos 2y_1 + (m_1 + m_3)\cosh 2y_2 \Big] \nonumber \\
 &\simeq& \frac{1}{4 m_{N_1}} (m_1 + m_3) e^{2|y_2|}  \quad\text{for}~ |y_2| \gg 1. 	
 \end{eqnarray}
In the limit $y_1 = 0,~ y_2 = 0$, we get $(R^\dagger R)_{11} = m_1/m_{N_1}$ and therefore restores the long-lived nature of $N_1$. 
For $N_2$ decay width, $(R^\dagger R)_{22} = m_2/m_{N_2}~ \forall~ y.$
For the case of $N_3$ decay width,  
\begin{eqnarray}
	(R^\dagger R)_{33} &=& \frac{1}{2m_{N_3}}\Big[ (-m_1 + m_3)\cos 2y_1 + (m_1 + m_3)\cosh 2y_2 \Big] \nonumber \\
	&\simeq& \frac{1}{4m_{N_3}} (m_1 + m_3)e^{2|y_2|}  \quad\text{for}~ |y_2| \gg 1. 
\end{eqnarray}
In the limit, $y_1 = 0,~ y_2 = 0$, we get $(R^\dagger R)_{33} = m_3/m_{N_3}$.

Finally, we repeat the same exercise with $y = z= 0$ and $x = x_1 + i x_2 \neq 0$. $(R^\dagger R)_{11} = m_1/m_{N_1}$ $\forall x$. For the decay width of $N_2$,
\begin{eqnarray}
	(R^\dagger R)_{22} &=& \frac{1}{2 m_{N_2}}\Big[(m_2 - m_3) \cos 2x_1 + (m_2 + m_3)\cosh 2 x_2 \Big] \nonumber \\
	&\simeq& \frac{1}{4m_{N_2}}(m_2 + m_3)e^{2|x_2|} \quad \text{for}~ |x_2| \gg 1. 
\end{eqnarray} 
In the limit, $x_1 = 0,~ x_2 = 0$, we get $(R^\dagger R)_{22} = m_2/m_{N_2}$ For the decay width of $N_3$, 
\begin{eqnarray}
	(R^\dagger R)_{33} &=& \frac{1}{2 m_{N_3}} \Big[ (-m_2 + m_3)\cos 2 x_1 + (m_2 + m_3)\cosh 2x_2 \Big] \nonumber \\
	&\simeq& \frac{1}{4m_{N_3}}(m_2 + m_3)e^{2 |x_2|} \quad \text{for}~ |x_2| \gg 1.
\end{eqnarray}
In the limit, $x_1 = 0,~ x_2 = 0$, we get $(R^\dagger R)_{33} = m_3/m_{N_3}$.

From these exercises, we can conclude that if $\mathcal{O} = \mathbb{I}$ then $N_1$ and $N_3$ can be long-lived in the case of NH and IH, respectively. However, if the $\mathcal{O} \neq \mathbb{I}$, then the HNL decay widths can be enhanced, and only in special cases we can have the long-lived scenarios:
\begin{eqnarray}
	\text{For} && x \neq 0, ~ y = z = 0, \quad N_1 \rightarrow \text{long-lived in NH} \nonumber \\
  \text{and} && z \neq 0, ~ x = y = 0, \quad N_3 \rightarrow \text{long-lived in IH}
\end{eqnarray}
In this paper we consider that $\mathcal{O} = \mathbb{I}$, \textcolor{black}{as a simple choice for making one of the HNL states long-lived.} 

\subsection{Case II: Minimal seesaw}

In the minimal seesaw, only two HNLs participate in the seesaw mechanism. Therefore in Eq.~[\ref{Eq: Neutrino mass term}] $m_D$ is $3\times 2$ matrix, $m_N$ is $2\times 2$ matrix and $R$ in Eq. \ref{Eq: Seesaw formula} is $3\times 2$ matrix. The light neutrino matrix $m_\nu$ and therefore $D_\nu =$ diag$(m_1, m_2, m_3)$ are $3\times 3$ matrices. However, $\sqrt{D_\nu}$ is a $3\times 2$ matrix. 

In NH, $m_{\text{lightest}} = m_1 =0$ and 
\begin{eqnarray}
\label{Eq: NH Dnu}
\sqrt{D^\text{NH}_\nu} = \left(\begin{array}{cc}
0 & 0 \\
\sqrt{m_2} & 0\\
0 & \sqrt{m_3}
\end{array}\right)
\end{eqnarray}
such that $D_\nu^{\text{NH}} = \sqrt{D_\nu^{\text{NH}}} \sqrt{D_\nu^{\text{NH}}}^T = \text{diag}(m_1=0, m_2, m_3)$. In IH, $m_{\text{lightest}} = m_3 = 0$ and
\begin{eqnarray}
\label{Eq: IH Dnu}
\sqrt{D^\text{IH}_\nu} = \left(\begin{array}{cc}
\sqrt{m_1} & 0\\
0 & \sqrt{m_2}\\
0 & 0 
\end{array}\right)
\end{eqnarray}
such that $D_\nu^{\text{IH}} = \sqrt{D_\nu^{\text{IH}}} \sqrt{D_\nu^{\text{IH}}}^T = \text{diag}(m_1, m_2, m_3 = 0)$.
In Eq.~[\ref{Eq: Cassas Ibarra}]
for the case of minimal seesaw, $\mathcal{O}$ is an $2\times 2$ orthogonal matrix  given by
\begin{eqnarray}
   \mathcal{O} = \left(\begin{array}{cc}
c_{z} & s_z\\
-s_z & c_z\\
\end{array}\right)
\label{EQ: Orthogonal matrix 2 X 2}
\end{eqnarray}
where $z=z_1 + iz_2$ is a complex parameter. Considering the general expression for the light-heavy neutrino mixing angle given in Eq.~[\ref{Eq: R matrix}], 
\begin{eqnarray}
    R^\dagger R = \sqrt{m_N^{-1}}\mathcal{O}^\dagger\sqrt{D_\nu}^T\sqrt{D_\nu}\mathcal{O}\sqrt{m_N^{-1}}.
\end{eqnarray}
In NH, 
\begin{eqnarray}
    \sqrt{D^{\text{NH}}_\nu}^T\sqrt{D^{\text{NH}}_\nu} = 
    \left(\begin{array}{cc}
m_2 & 0\\
0 & m_3
\end{array}\right),
\end{eqnarray}
and similarly in IH, 
\begin{eqnarray}
    \sqrt{D^{\text{IH}}_\nu}^T\sqrt{D^{\text{IH}}_\nu} = 
    \left(\begin{array}{cc}
m_1 & 0\\
0 & m_2
\end{array}\right).
\end{eqnarray}
Setting the orthogonal matrix $\mathcal{O}=\mathbb{I}$, we get 
\begin{eqnarray}
    R^\dagger R = m_N^{-1} \sqrt{D_\nu}^T\sqrt{D_\nu},
\end{eqnarray}
so that in NH we get $(R^\dagger R)^{\text{NH}}_{11}= m_2/m_{N_1}$ and $(R^\dagger R)^{\text{NH}}_{22}= m_3/m_{N_2}$. Whereas, for IH $(R^\dagger R)^{\text{IH}}_{11}= m_1/m_{N_1}$ and $(R^\dagger R)^{\text{IH}}_{22}= m_2/m_{N_2}$. Since $(R^\dagger R)_{ii}$ in the minimal seesaw is not proportional to the lightest neutrino, the HNLs in this scenario are not long-lived, compared to the HNLs in the standard seesaw. 

Now, repeating the same exercise, but with  $\mathcal{O}\neq\mathbb{I}$, for $z_1 = 0$ and $z_2 \neq 0$, in NH we get 
\begin{eqnarray}
    (R^\dagger R)^{\text{NH}}_{11} = \frac{1}{m_{N_1}}[m_2 \cosh^2 z_2 + m_3 \sinh^2 z_2], \nonumber \\
    (R^\dagger R)^{\text{NH}}_{22} = \frac{1}{m_{N_2}}[m_3 \cosh^2 z_2 + m_2 \sinh^2 z_2],    
\end{eqnarray}
and in IH we get 
\begin{eqnarray}
    (R^\dagger R)^{\text{IH}}_{11} = \frac{1}{m_{N_1}}[m_1 \cosh^2 z_2 + m_2 \sinh^2 z_2], \nonumber \\
    (R^\dagger R)^{\text{IH}}_{22} = \frac{1}{m_{N_2}}[m_2 \cosh^2 z_2 + m_1 \sinh^2 z_2],    
\end{eqnarray}
 Similar to the standard seesaw, the decay widths of the HNLs in the minimal seesaw can also be enhanced for $|z_2|\gg 1$. Since our goal is to study long-lived HNLs in the LHC, we are restricted only in the context of the standard seesaw mechanism with $\mathcal{O}=\mathbb{I}$.

\subsection{Neutrinophilic Charged Higgs}
The $H^\pm$ decay to the HNL states is given by
\textcolor{black}{
\begin{eqnarray}
	\Gamma(H^+ \to \ell^+_{\alpha} N_i) = \frac{|{m_{D}}_{\alpha i}|^2}{8 \pi v_2^2}m_{H^\pm} \Big( 1 - \frac{m^2_{N_i}}{m_{H^\pm}^2} \Big)^2,
	\label{Eq: Hpm decay width} 
\end{eqnarray}
}
where $v_2$ is the VEV of the neutrinophilic second Higgs doublet and like before, the indices $\alpha$ and $i$ are the SM lepton flavor and HNL mass eigenstate indices. Since, the $H^\pm$ comes dominantly from the second Higgs doublet, which is neutrinophilic, $H^\pm$ decays to the HNL states by almost $100\%$.    

Summing over the charged lepton states in Eq.~[\ref{Eq: Hpm decay width}], we get the factor
\begin{eqnarray}
    \sum_{\alpha=e,\mu,\tau} |{m_{D}}_{\alpha i}|^2 = (m_D^\dagger m_D)_{ii}, 
\end{eqnarray}
where
\begin{eqnarray}
\label{Eq: mD dag mD}
    m_D^\dagger m_D = \sqrt{m_N} \mathcal{O}^\dagger D_\nu \mathcal{O}\sqrt{m_N}, 
\end{eqnarray}
and it is proportional to the HNL mass unlike the $R^\dagger R$ in Eq.~[\ref{Eq: RdR}] which is inversely proportional to it. Therefore, all the formulas for $(R^\dagger R)_{ii}$ in Sec.~\ref{Sec: Case I: Standard seesaw} hold true for the factor ${(m_D^\dagger m_D)}_{ii}$ with HNL mass $m_{N_i}$ in the numerator. Thus, $H^\pm$ is not long-lived compared to the HNLs in the standard seesaw.       

\section{Phenomenology}
\label{Sec: Phenomenology}

In this section, we show the phenomenology of the simple setup of the neutrinophilic $H^\pm$ as discussed in Sec.~\ref{Sec:Simple Setup}. We can have long-lived HNL states only for the standard type-I seesaw where all the three RHNs participate in the seesaw mechanism. In NH $N_1$ is long-lived, whereas in IH $N_3$ is long-lived. We will discuss the pair production of $H^\pm$ at the LHC, the decays of $H^\pm$ to the HNL states, followed by the decays of the HNL states. \textcolor{black}{For simplicity, we assume a degenerate mass spectrum for the extra Higgs bosons ($m_{H^\pm}\simeq m_{H} \simeq m_{A}$). We further assume that the long-lived HNL states  i.e. $N_1$ in NH and $N_3$ in IH to be lighter than  $H^\pm$, while the remaining HNL states to be heavier than $H^\pm$.} In other words, the decays of $H^\pm$ in NH and IH are $ \ell_\alpha N_1$ and $ \ell_\alpha N_3$, respectively, where $\ell_\alpha = e,\mu, \tau$. The long-lived HNLs $N_1$ or $N_3$, once produced, decays away from their production points into $\ell_\alpha W$ and $\nu_\alpha Z$. Before discussing these phenomenologies, we present the limits on the VEV of the neutrinophilic Higgs doublet $v_2$, from the lepton flavor violation (LFV) process such as $\mu \to e\gamma$. 

\subsection{LFV: $\mu \to e \gamma$}
\label{Sec: LFV: mu to e gamma}

\begin{figure}[t!]
\includegraphics[width=8.cm]{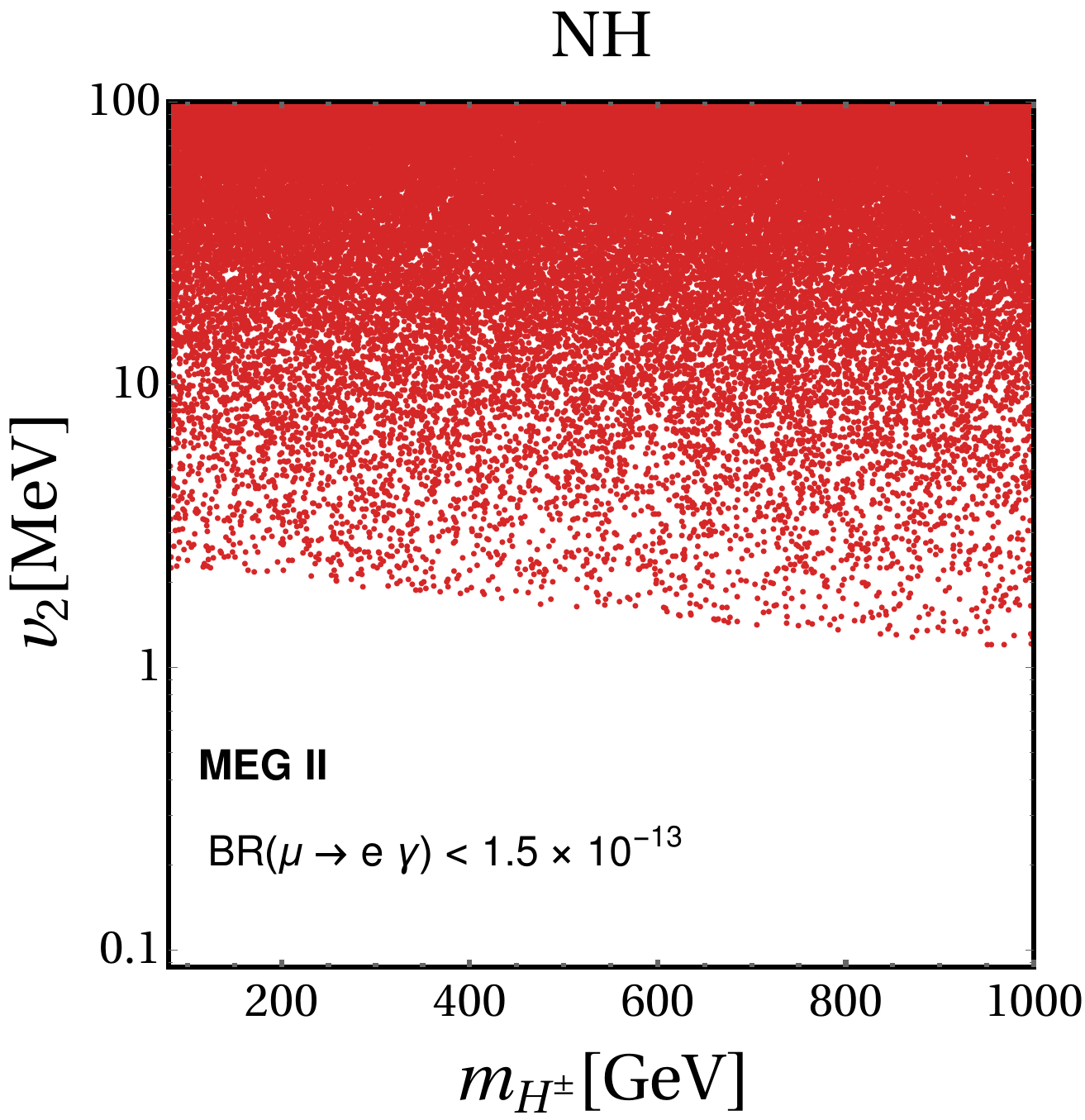}
\includegraphics[width=8.cm]{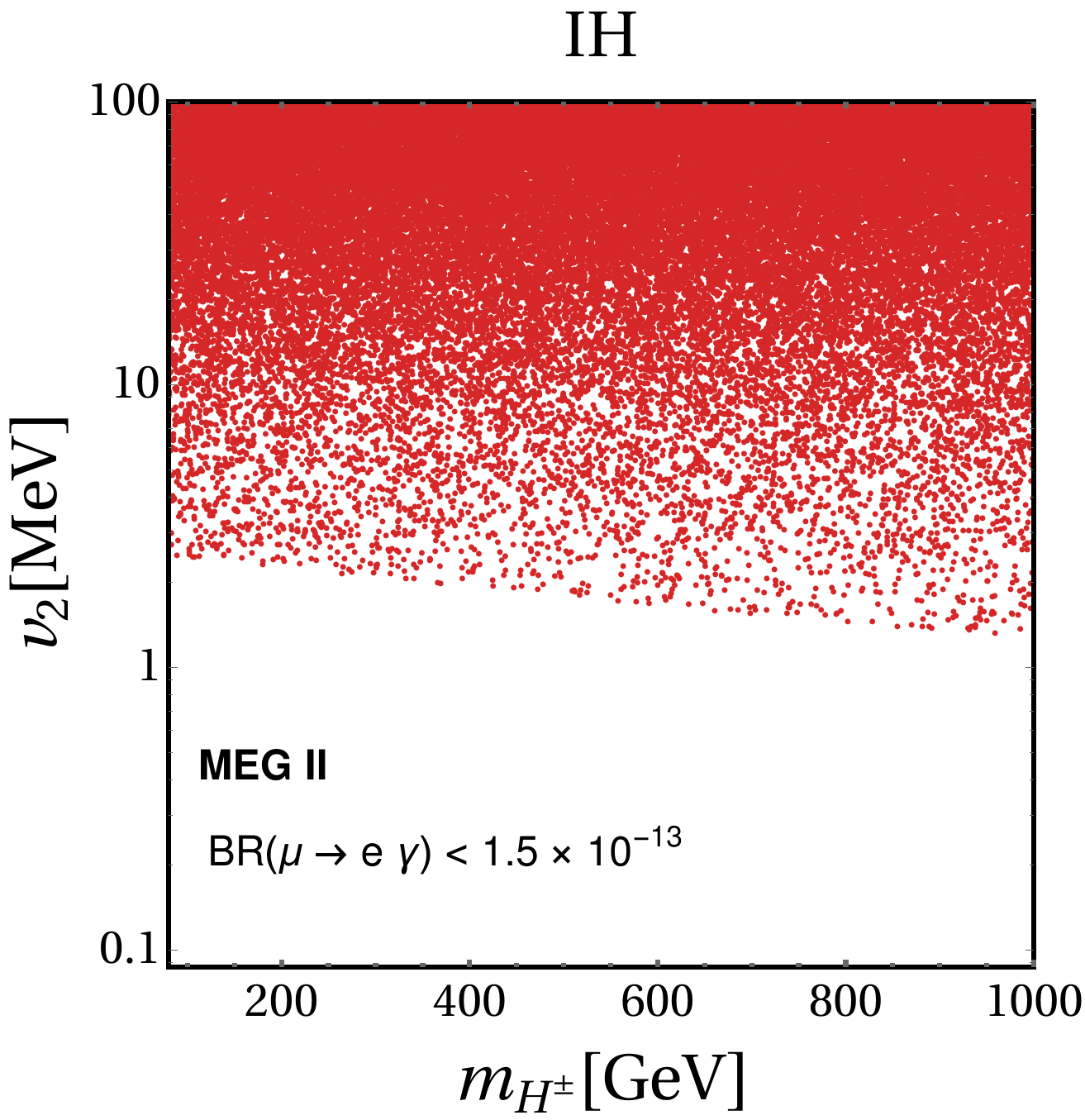}
\caption{Limits on the VEV  of the neutrinophilic second Higgs doublet $v_2$ from the $\mu \to e \gamma$ constraint. Left: Normal hierarchy (NH) scenario. Right: Inverted hierarchy (NH) scenario. }
\label{Fig: LFV mu to e gamma plots}
\end{figure}

The most stringent LFV constraint comes from the radiative decay $\mu \to e\gamma$. The current limit on the branching ratio from the MEG experiment is given as $\text{BR}(\mu \to e \gamma) < 1.5\times 10^{-13}$ \cite{MEGII:2025gzr}. The branching ratio of the process $\mu \to e \gamma$ in the standard type-I seesaw is given by \cite{He:2002pva, Ma:2001mr}
\begin{eqnarray}
\label{Eq: mu to e gamma}
      \text{BR}(\mu \to e\gamma) \simeq \frac{3\alpha_e}{32 \pi}\Bigg|\sum_{i=1}^3 R^*_{1i} R_{2i} F_1\Big(\frac{m^2_{N_i}}{m^2_W}\Big)\Bigg|^2 + \frac{3\alpha_e}{16 \pi G_F^2 m^4_{H^\pm} v_2^4}\Bigg|\sum_i^3 {m^*_D}_{1i}{m_D}_{2i} F_2\Big(\frac{m^2_{N_i}}{m^2_{H^\pm}}\Big)\Bigg|^2, 
\end{eqnarray}
where the loop functions $F_{1,2}(x)$ are 
\begin{eqnarray}
    F_1(x) &=& \frac{10-43x+78x^2 -49 x^3 + 18x^3\log x + 4 x^4}{6(1-x)^4}, \nonumber \\
    F_2(x) &=& \frac{1-6x+3x^2 + 2x^3 - 6x^2\log x}{6(1-x)^4}.
\end{eqnarray}
The first term in Eq.~[\ref{Eq: mu to e gamma}] comes from the HNL and $W$ boson loop, whereas the second term comes from the HNL and $H^\pm$ loop. The first term is suppressed in the regime of our interest, where we considered $\mathcal{O}=\mathbb{I}$ and the HNL states can be made long-lived. However, the contribution of the first term can be significantly enhanced in the scenarios where $\mathcal{O}\neq \mathbb{I}$~\cite{Morisi:2024yxi}. Since we focus on the case with $\mathcal{O}=\mathbb{I}$, the first term in Eq.~[\ref{Eq: mu to e gamma}] is negligible compared to the second term. 

To get an estimate of the VEV of the neutrinophilic Higgs doublet from the $\mu \to e\gamma$ constraint, we perform a parameter scan for $10^{-7}\leq m_{\text{lightest}} \leq 10^{-1}$ eV, $ 80\leq m_{H^\pm} \leq 1000$ GeV, $m_{N_1} < m_{H^\pm}$ in NH and $m_{N_3} < m_{H^\pm}$ in IH,  while fixing the other HNL states like $N_2, N_3$ as 1 TeV in NH and $N_1, N_2$ as 1 TeV in IH.  In Fig.~\ref{Fig: LFV mu to e gamma plots} we showed the allowed regions for the NH and IH scenarios. We see that of $v_2 \lesssim \mathcal{O}(1)$ MeV is excluded. For the rest of our analysis, we set  $v_2$ to be 10 MeV.

\subsection{HNL Production and Decays}
\label{Sec: HNL Production and Decays}
\begin{figure}[t!]
\includegraphics[width=10cm]{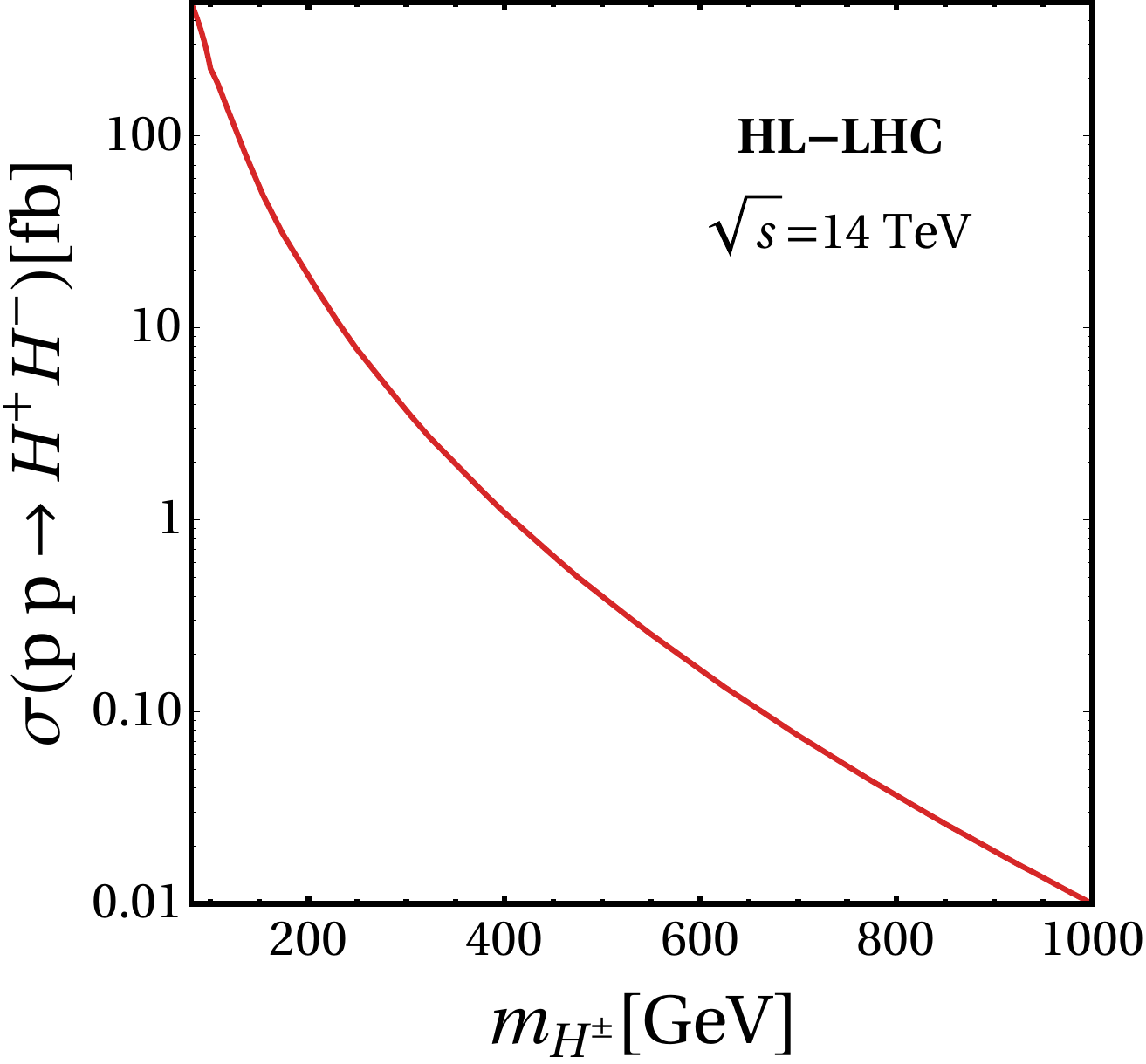}
\caption{Charged Higgs pair production cross section at the 14 TeV LHC.}
\label{Fig: Hpm pair creation}
\end{figure}

The long-lived HNL states $N_1$ in NH and $N_3$ in IH can be produced from the decays of the non-SM Higgs bosons which come predominantly from the neutrinophilic second Higgs doublet. In this paper we consider a neutrinophilic $H^\pm$, which is pair produced in the LHC via the $\gamma/Z$ mediated $s$-channel EW process.  The pair production cross section of $H^\pm$ at the 14 TeV HL-LHC is shown in Fig.~\ref{Fig: Hpm pair creation}. The cross section is computed using \texttt{MadGraph5\_aMC@NLO-3.5.3} \cite{Alwall:2011uj} with the \texttt{NNPDF31\_lo\_as\_0118} parton distribution function (PDF) set \cite{NNPDF:2017mvq}, interfaced with a \texttt{Universal FeynRules Output} (UFO) \cite{Degrande:2011ua} model from \texttt{FeynRules-2.3} \cite{Alloul:2013bka}. From Fig.~\ref{Fig: Hpm pair creation}, we can see that the $H^\pm$ pair production cross section falls sharply with the $H^\pm$ mass and for $m_{H^\pm}\gtrsim 400$ GeV, the cross section falls below 1 fb.  

\begin{figure}[t!]
\includegraphics[width=8cm]{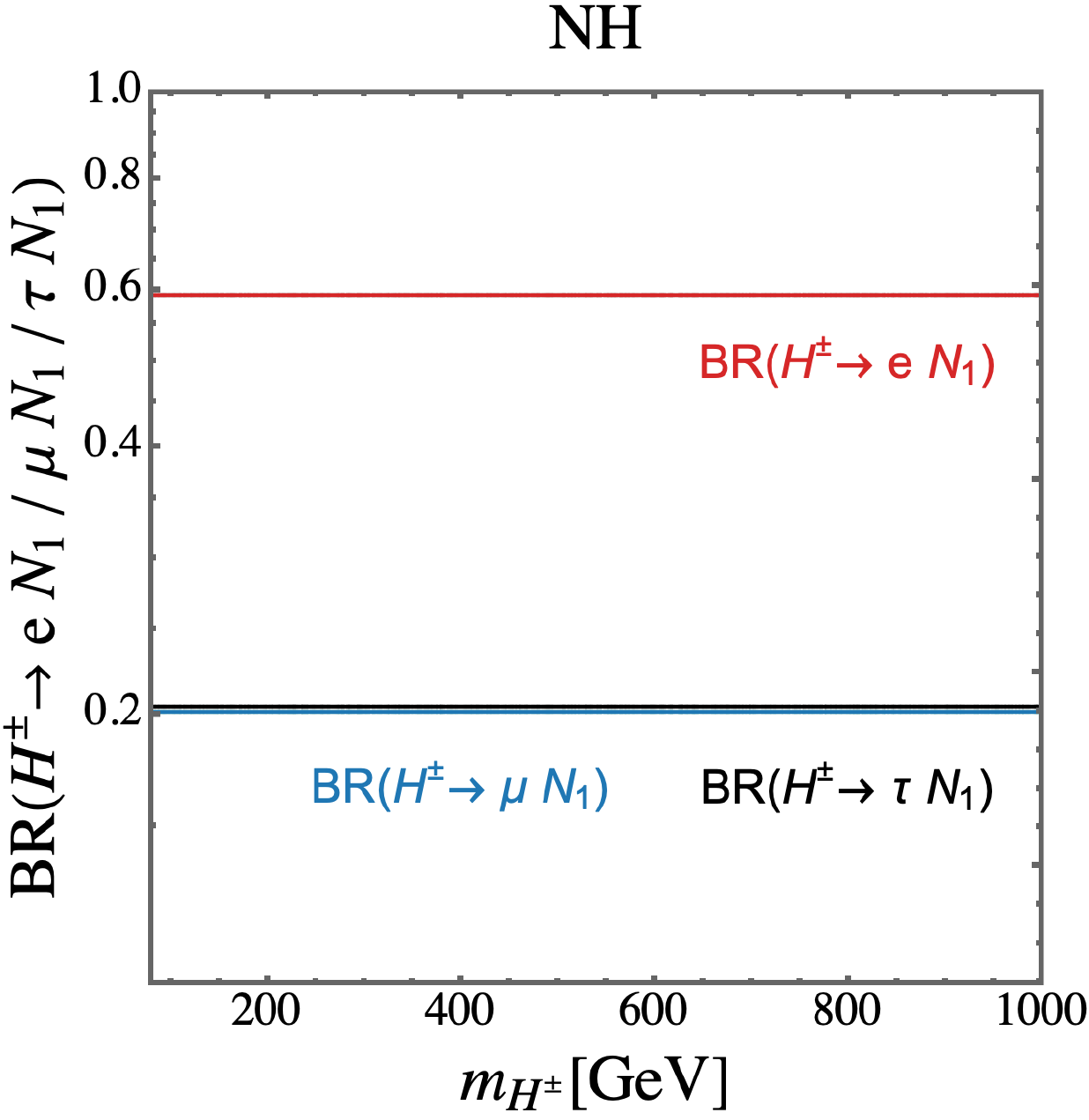}
\includegraphics[width=8cm]{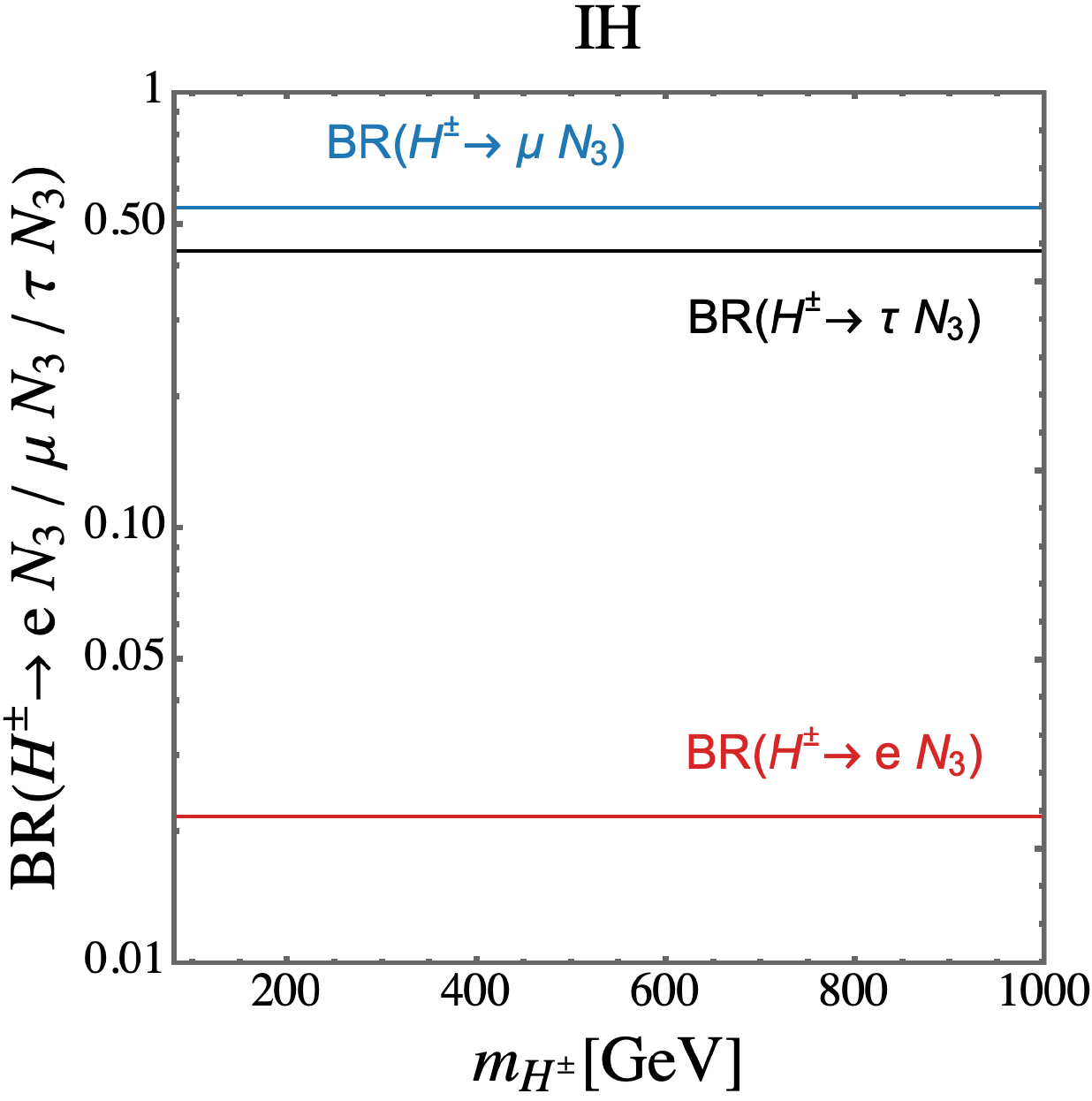}
\includegraphics[width=8cm]{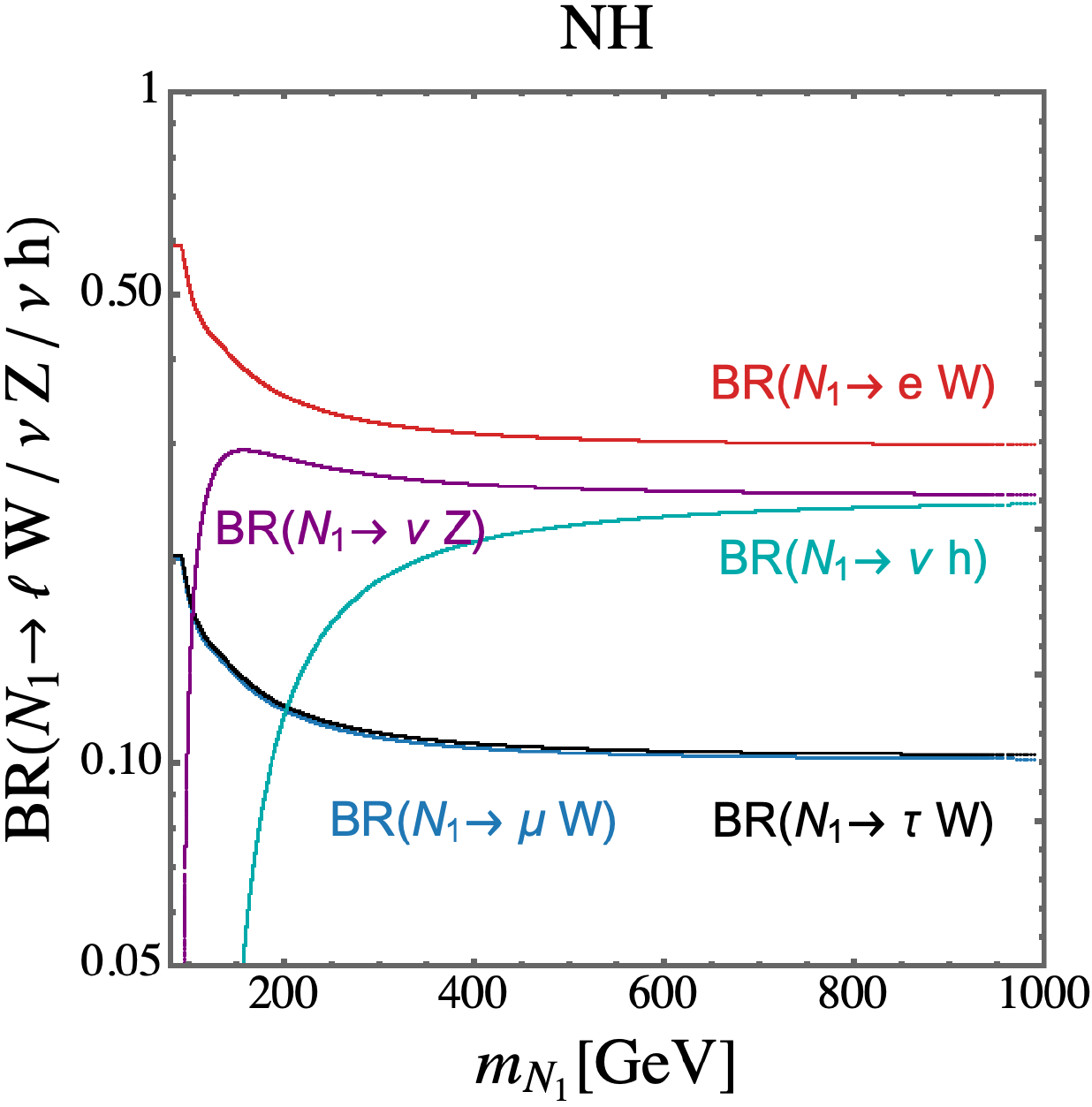}
\includegraphics[width=8cm]{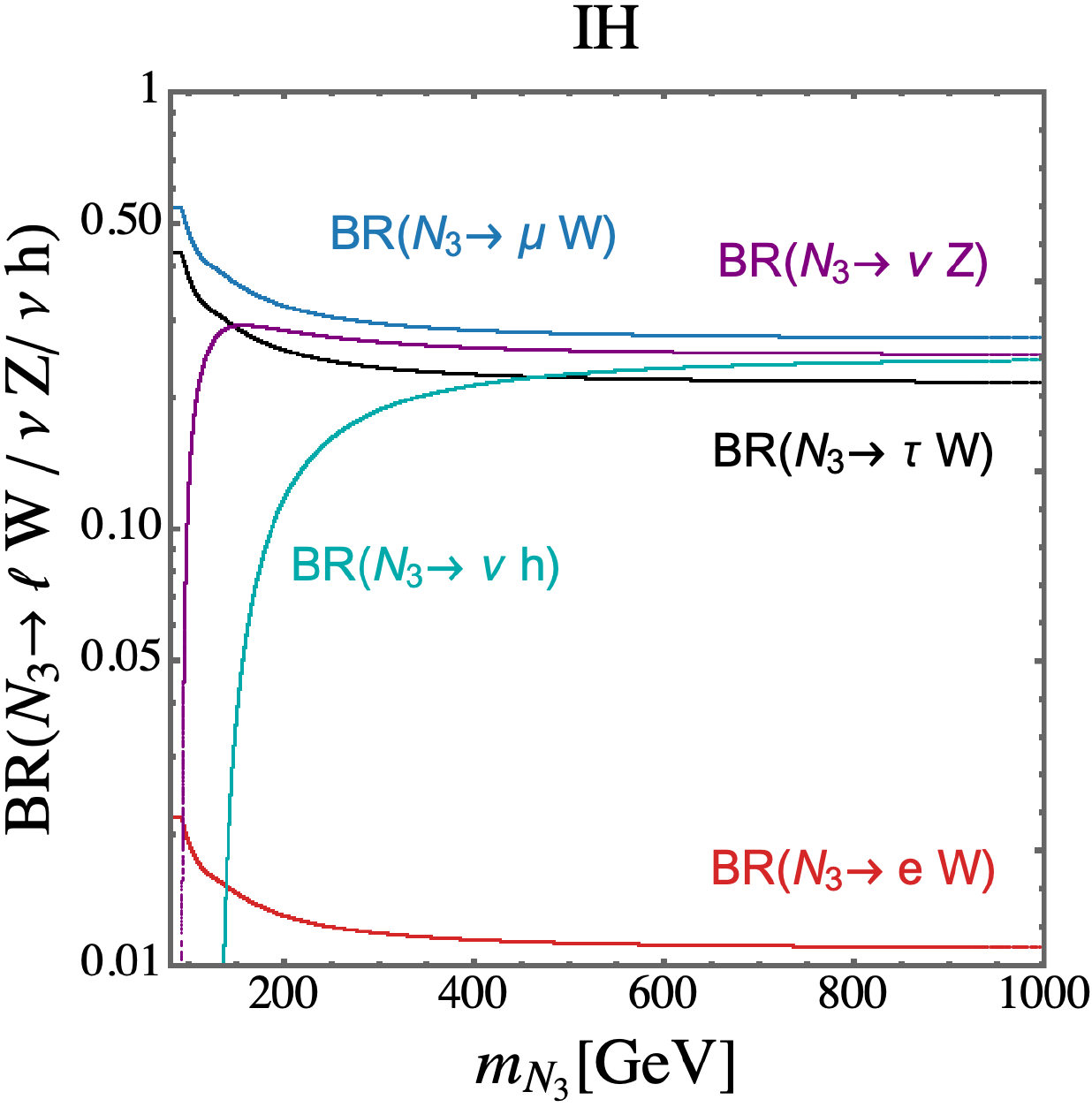}
\caption{Top: $H^\pm$ branching ratios to the long-lived HNLs in normal hierarchy (NH) (left) and inverted hierarchy (IH) (right). Bottom: Long-lived HNL branching ratios in NH (left) and IH (right).}
\label{Fig: BR plots}
\end{figure}

The $H^\pm$ pair decays promptly to the HNL states as given in Eq.~[\ref{Eq: Hpm decay width}]. Since the long-lived HNL states $N_1$ and $N_3$ are considered lighter than $H^\pm$ in NH and in IH, respectively, $H^\pm$ decays only to the long-lived HNL states along with charged SM leptons. The $H^\pm$ branching ratios to $\ell_\alpha N_1$ and $\ell_\alpha N_3$ for NH and IH are shown in the top panel in Fig.~\ref{Fig: BR plots}. We can see that in NH $H^\pm$ decays dominantly to $e N_1$, while in IH $H^\pm$ decays dominantly to $\mu N_3$ followed by $\tau N_3$. In the lower panel of Fig.~\ref{Fig: BR plots}, we show the branching ratios of the long-lived HNL states. \textcolor{black}{The partial decay widths of the HNL states are given in Eq.~[\ref{Eq: HNL decay widths}]. In NH  $N_1$ decays dominantly to $eW$, while in IH $N_3$ decays dominantly to $\mu W$. The branching ratios of $N_i \to \nu Z/\nu h$ include the three flavors of light neutrinos.}          

\begin{figure}[t!]
\includegraphics[width=5.25cm]{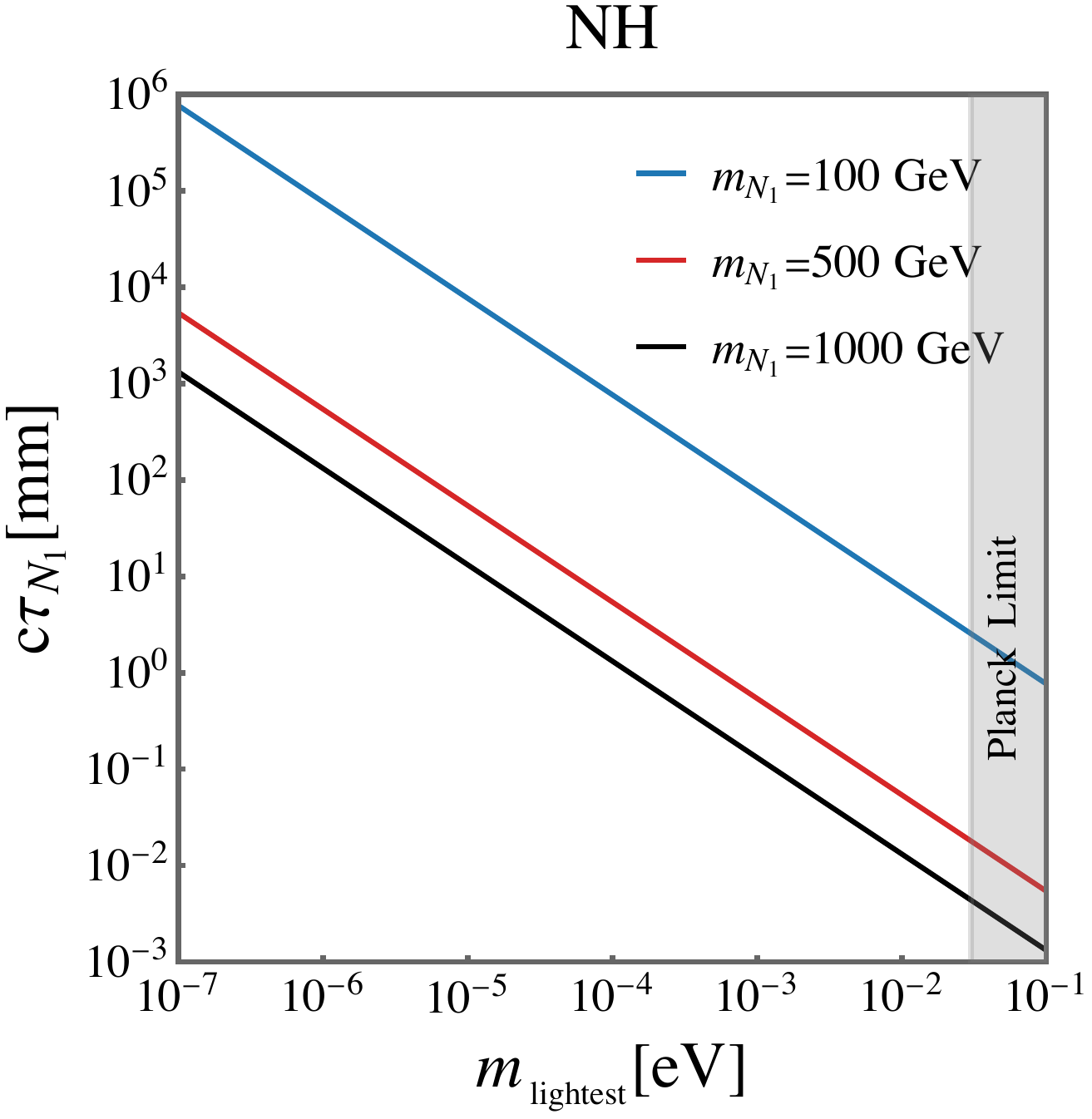}
\includegraphics[width=5.25cm]{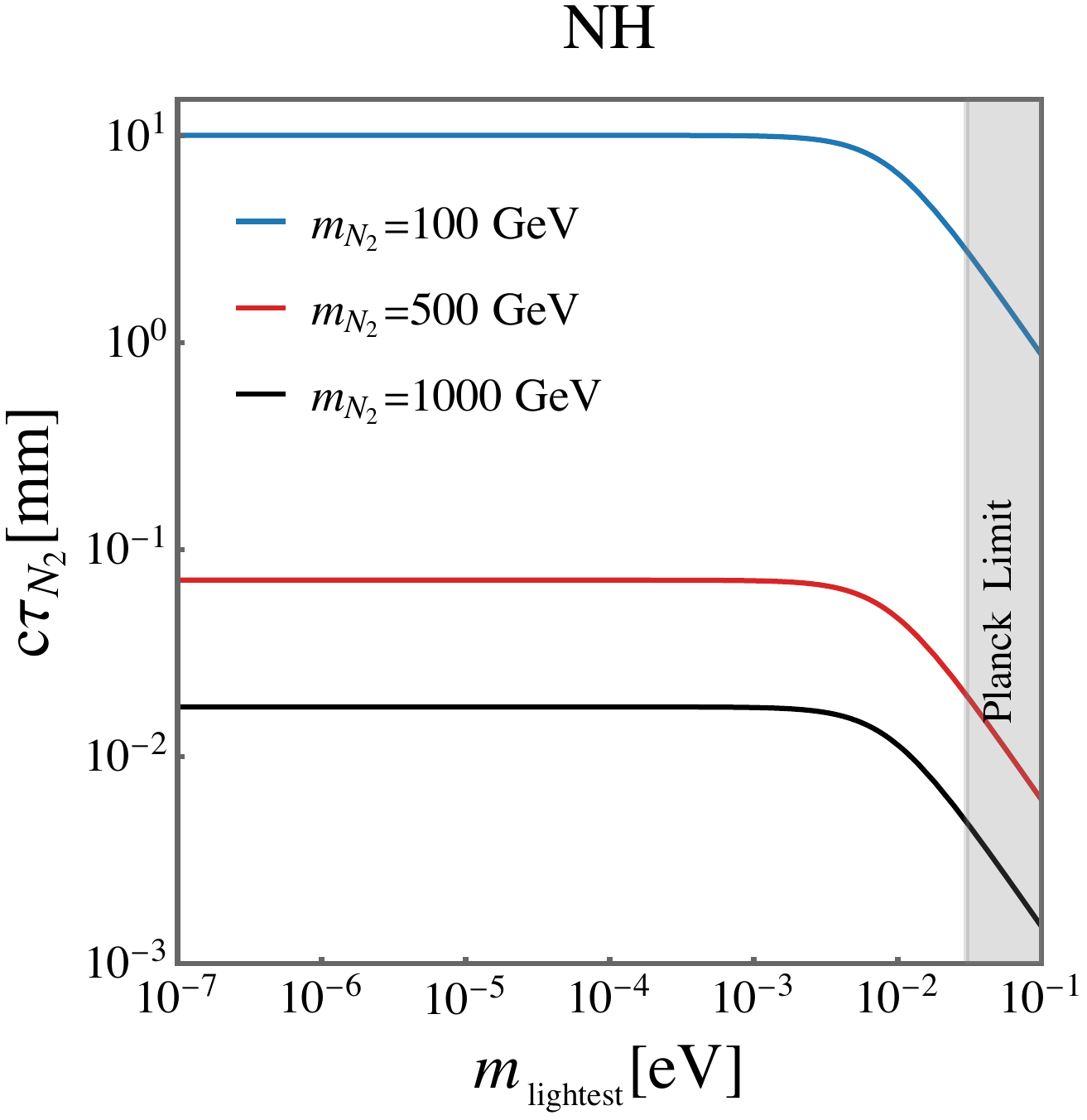}
\includegraphics[width=5.25cm]{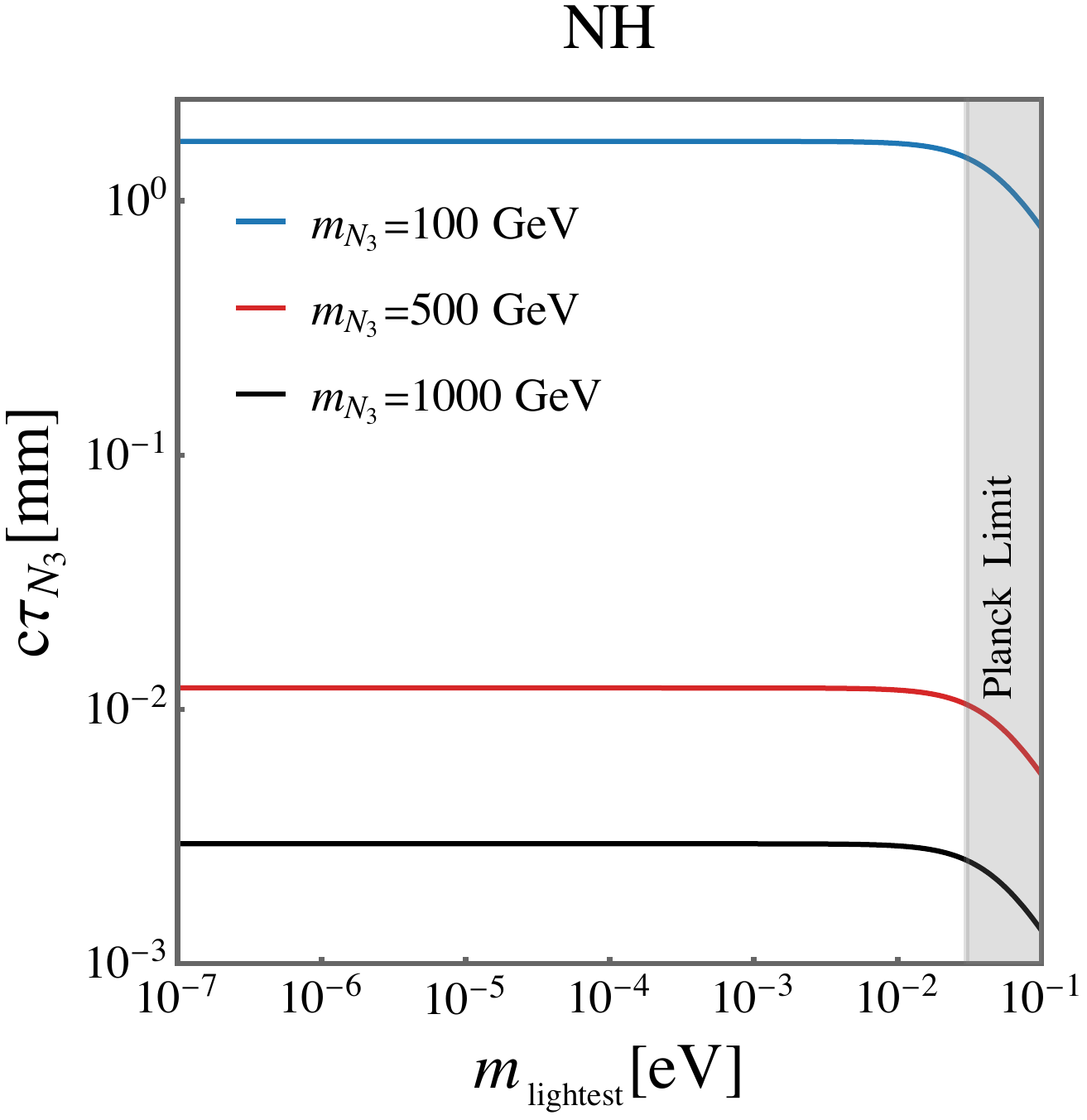}
\includegraphics[width=5.25cm]{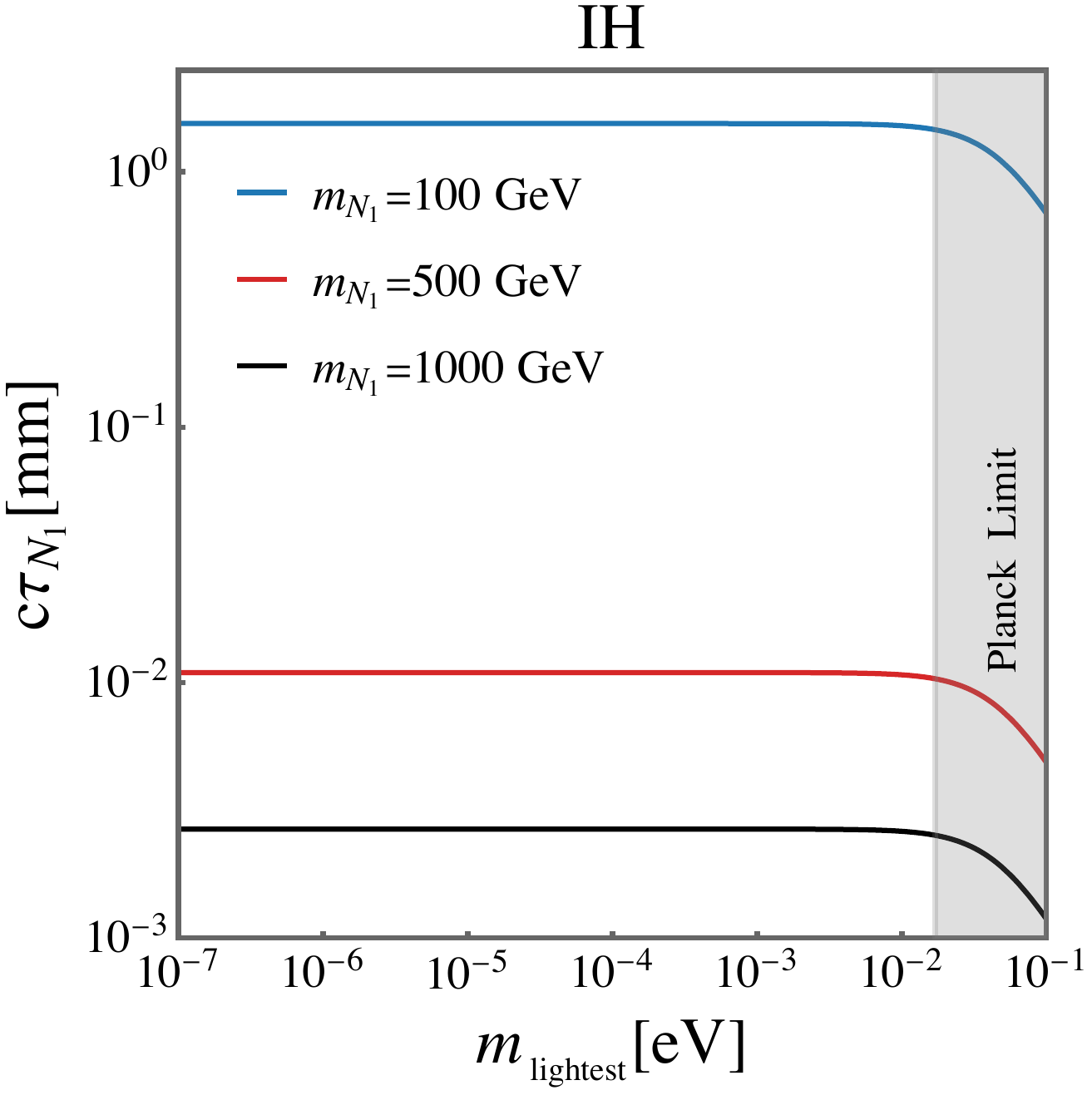}
\includegraphics[width=5.25cm]{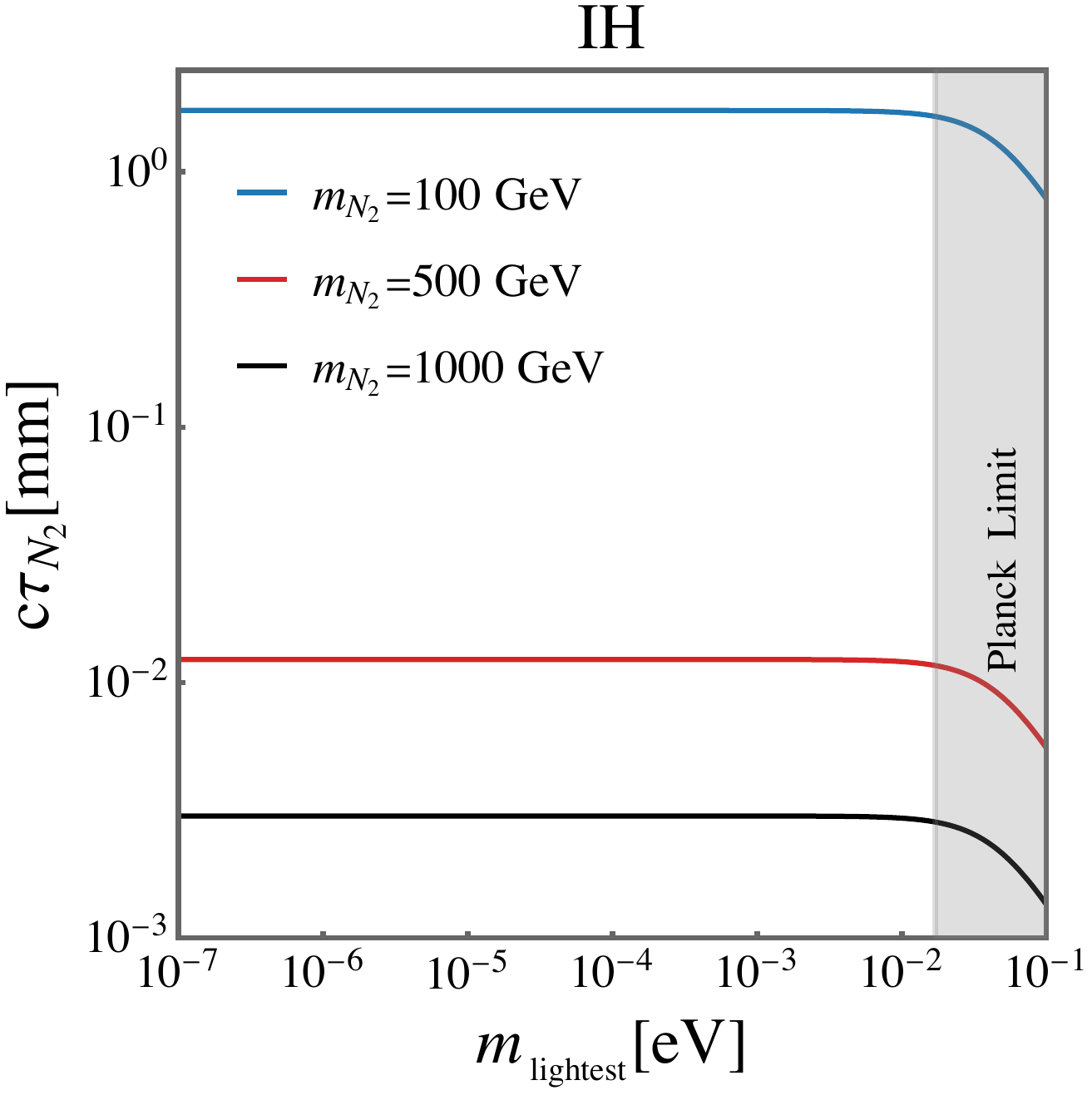}
\includegraphics[width=5.25cm]{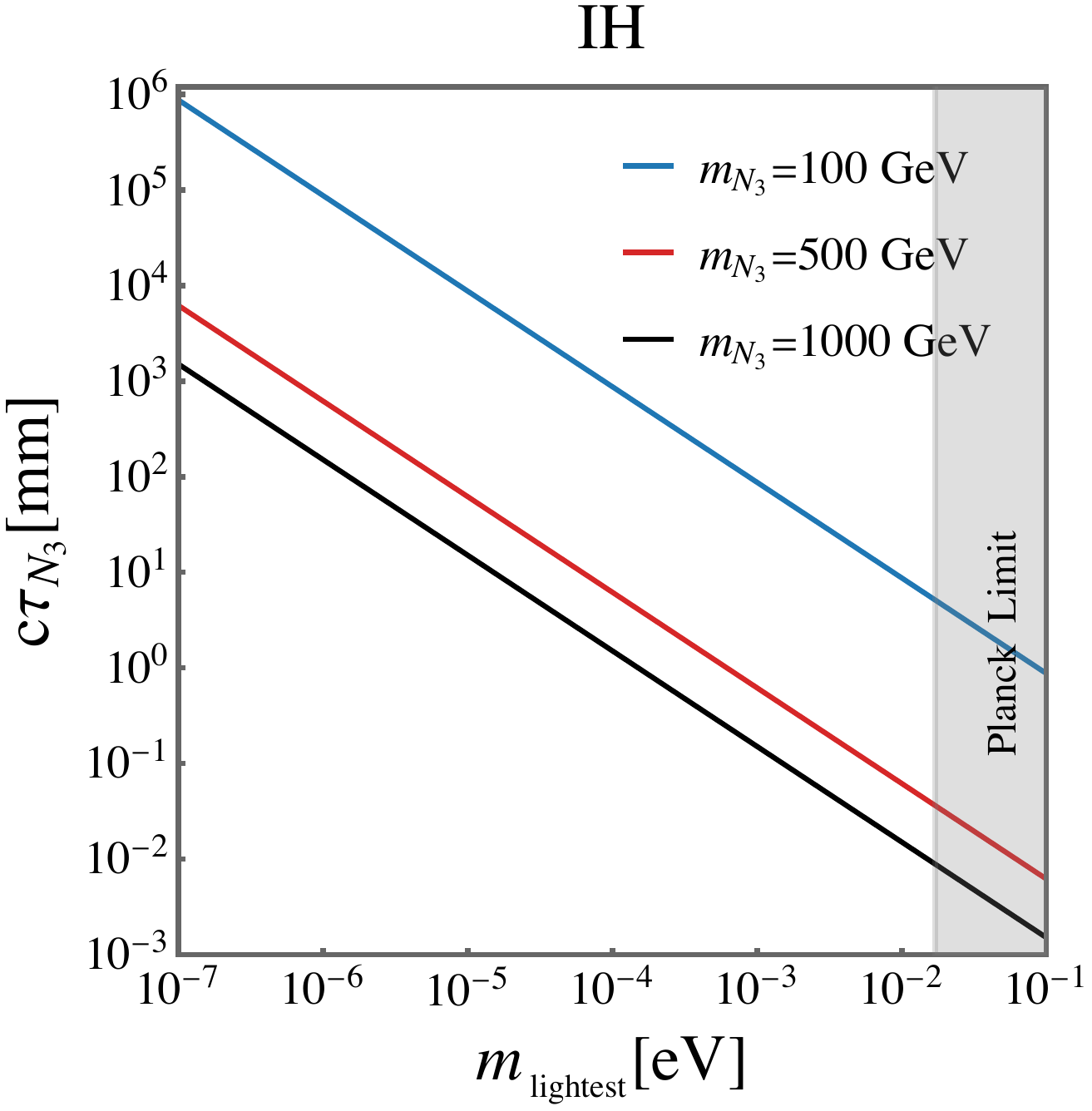}
\caption{Top: Proper decay lengths of the HNL states in normal hierarchy (NH). Bottom: Proper decay lengths of the HNL states in inverted hierarchy (IH). The vertical shaded region is excluded by the Planck upper limit on the sum of neutrino masses.}
\label{Fig: HNL decay length}
\end{figure}

Finally, in Fig.~\ref{Fig: HNL decay length} we present the proper decay lengths of the HNL states for the NH (top panel) and IH (bottom panel) scenarios. In the NH case, $N_1$ is the long-lived HNL state, whereas in the IH case $N_3$ is the long-lived one. We observe that the decay length of the long-lived HNL increases as the mass of the lightest neutrino decreases. Such an enhancement of the decay length does not occur for the other two HNL states in either the NH or IH scenarios. To illustrate this behavior, we considered three representative HNL masses 100 GeV 500 GeV, and 1000 GeV.


\section{Collider Simulation and Analysis}
\label{Sec: Collider Simulation and Analysis}

\begin{figure}[t!]
	\includegraphics[width=15cm]{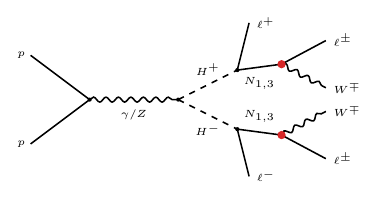}
	\caption{Feynman diagram for the production of long-lived HNLs ($N_1$ in the normal hierarchy (NH) and $N_3$ in the inverted hierarchy (IH)) from $H^\pm$ pair production. The displaced vertices (DVs) associated with the long-lived HNL decays are shown in red.}
	\label{Fig: Feynman Diagram}
\end{figure}

At the LHC, the $H^\pm$ is pair produced via the EW process and decays to the long-lived HNL states $N_1$ in NH and $N_3$ in IH. As shown in Fig.~\ref{Fig: BR plots}, the long-lived HNL states decays dominantly to the $\ell_\alpha W$ mode. The Feynman diagram for the dominant process under consideration, as shown in Fig.~\ref{Fig: Feynman Diagram}, is given by\footnote{\textcolor{black}{Several other interesting LHC signatures are also worth investigating, including
	$pp\to H^\pm H/A, H^\pm\to \ell^\pm N, H/A\to \nu N,$
	as well as
	$pp\to HA, H/A\to \nu N.$}}
\begin{eqnarray}
    p p \to H^+H^- \to (\ell^+ N_{1,3} \to \ell^+ \ell^\pm W^\mp)(\ell^- N_{1,3} \to \ell^- \ell^\pm W^\mp).
    \label{Eq: Signal process}
\end{eqnarray}
The displaced vertices (DVs) associated with the long-lived HNL states are shown in red. The $W$ bosons are allowed to decay into the hadronic mode $W^\pm \to jj$, leading to a final state consisting of $4\ell4j$. Considering the dominant decay modes of $H^\pm$ and the long-lived HNL states as shown in Fig.~\ref{Fig: BR plots}, we can have the $4e4j$ (muon veto) and $4\mu4j$ (electron veto) modes indicating the NH and IH of neutrino masses. Additionally, if three of the 4-leptons are identified to have the same sign, we can see a lepton number violation (LNV) by two units. The LNV by two units would be a clear indication of the Majorana nature of neutrinos. 

The signal events at the parton-level are generated with \texttt{MadGraph5}, interfaced to \texttt{Pythia-8.3} \cite{Bierlich:2022pfr} for parton showering and hadronization. The decays of unstable SM particles, including $\tau$ leptons and $W$ bosons, are handled in \texttt{Pythia}. The detector effect at the HL-LHC is simulated using \texttt{Delphes-3.5.0} \cite{deFavereau:2013fsa} with the \texttt{delphes\_card\_HLLHC} configuration card. Jets are clustered using the anti-$k_T$ algorithm~\cite{Cacciari:2008gp}, as implemented in \texttt{FastJet-3.3.4}~\cite{Cacciari:2011ma}, with a jet radius parameter $R = 0.4$. The proper decay time $\tau$ is sampled from the exponential probability distribution 
\begin{equation}
   P(\tau)d\tau =  \Gamma_{N_i} e^{-\Gamma_{N_i}\tau}d\tau,
\end{equation}
where $\Gamma_{N_i}$ is the total width of the HNL states $N_1$ and $N_3$ in the NH and IH, respectively. The decay length in the laboratory (LHC) frame is then obtained via the boost relation
\begin{eqnarray}
    L_{\text{lab}} = \beta\gamma c\tau,
\end{eqnarray}
where $\beta = |\vec{p}|/E_{N_i}$ is the velocity in units of the speed of light $c$ and $\gamma = E_{N_i}/m_{N_i}$ is the Lorentz factor in the laboratory frame.
We set the run-card parameter \texttt{time\_of\_flight} in \texttt{MadGraph5} as zero to activate the computation and storage of \texttt{vtim} $=c\tau$ information.

An appropriate reconstruction of the displaced vertex within the \texttt{Delphes} framework is beyond the scope of this paper. We design our search strategy to target displaced leptons without imposing an explicit displaced vertex reconstruction requirement, thereby covering a broader class of event topologies. Searches have been conducted that select events with an electron and muon which are displaced, e.g. Refs. \cite{CMS:2014xnn, CMS:2016isf}. The searches at 8 TeV \cite{CMS:2014xnn} and 13 TeV \cite{CMS:2016isf} differ primarily in the requirement of the lepton transverse momenta. At 8 TeV both leptons satisfy $p_T(e,\mu) > 25$ GeV, whereas at 13 TeV $p_T(e)> 42$ GeV and $p_T(\mu) >40$ GeV. The displacement of leptons is characterized by the unsigned transverse impact parameter $|d_0|$, defined as the distance of the closest approach of the lepton track to the primary vertex in the azimuthal plane. Based on the values of $|d_0|$, they defined three mutually exclusive signal regions and obtained electron and muon $|d_0|$ dependent efficiencies. Polynomial parameterizations of the efficiencies for $|d_0| < 20~\text{mm}$ are provided in Ref.~\cite{Chang:2018gqh}, while for $|d_0| > 20~\text{mm}$ a linear extrapolation, vanishing at $|d_0| = 100~\text{mm}$, is adopted as given in Ref.~\cite{Araz:2021akd}. The $|d_0|$ efficiencies for electrons and muons for $|d_0| < 20~\text{mm}$ are given as:
\begin{eqnarray}
\label{Eq: d0 efficiencies_1}
\epsilon_e(x) &=&
0.924921
- 0.0917957\,x
+ 0.00522007\,x^2
+ 0.00287189\,x^3
- 0.00049321\,x^4 \nonumber \\
&+& 2.72756\times10^{-5}\,x^5 
- 5.06107\times10^{-7}\,x^6, \nonumber \\
\epsilon_\mu(x) &=&
0.99067
- 0.0271852\,x
+ 0.00743217\,x^2
- 0.000611108\,x^3
- 2.60292\times10^{-5}\,x^4 \nonumber \\
&+& 4.23266\times10^{-6}\,x^5
- 1.11279\times10^{-7}\,x^6 ,
\end{eqnarray}
where $x= |d_0|/\text{mm}$. The efficiencies for $20~\text{mm}<|d_0| \leq 100~\text{mm}$ read as
\begin{eqnarray}
\label{Eq: d0 efficiencies_2}
    \epsilon_{e}(x) &=& 0.15 -0.001875(x-20), \nonumber \\
    \epsilon_{\mu}(x) &=& 0.8 -0.01(x-20),
\end{eqnarray}
vanishing for $|d_0|$ larger than $100$ mm.

In our analysis, as mentioned above, we used the  \texttt{delphes\_card\_HLLHC} configuration card, with its default tracking efficiencies, lepton isolation and identification criteria. \texttt{Delphes} provides information on the lepton transverse component of the impact parameter $|d_0|$ and the longitudinal component along the beam axis  $|d_z|$, estimated from \texttt{vtim} computed in \texttt{MadGraph5}. It also applies a veto to leptons produced outside the tracking volume. Since \texttt{Delphes} does not include the tracking efficiency as a function of lepton displacement, we incorporate an additional correction at the analysis level to account for the reduced reconstruction efficiency as a function of $|d_0|$, as given in Eqs.~[\ref{Eq: d0 efficiencies_1}] and [\ref{Eq: d0 efficiencies_2}]. We apply a relative displacement correction factor defined as
\begin{eqnarray}
    f_\ell(x) = \frac{\epsilon_{\ell}(x)}{\epsilon_{\ell}(x = 0)}, \quad\quad \ell=e,\mu,
\end{eqnarray}
where, like before $x= |d_0|/\text{mm}$. The event weight is multiplied by the product of the relative displacement correction factor for each selected displaced lepton. By construction, $f_\ell(0) = 1$ ensures that only the additional suppression due to displacement is applied.

The process considered in Eq.~[\ref{Eq: Signal process}] has two DVs associated with the two long-lived HNL states, as shown in Fig.~\ref{Fig: Feynman Diagram} in red. The tagging efficiency of two DVs is much smaller than the tagging efficiency of at least one DV, as can be understood from the efficiencies $|d_0|$ associated with the displaced leptons. Therefore, one DV signature would have a wider sensitive region than the two DV signature. Hence, in our analysis we have studied the signatures of at least one DV (1DV) as well as two DV (2DV). In both cases, we require at least one prompt lepton from the interaction point that one can trigger on.
The lepton isolation criteria are Iso$(e)< 0.1$ 
and Iso$(\mu) < 0.2$, where isolation is defined as the sum of the $p_T$ of all other reconstructed particles within a cone of $R= 0.3$ around the lepton, divided by the $p_T$ of the lepton.
Below we mention the selection criteria adopted for the 1DV and 2DV signatures based on the signal topologies. 

\begin{enumerate}[label={}]
    \item \textbf{1DV selection criteria:}
    \begin{enumerate}
        \item \textbf{Preselection:} We require at least one prompt lepton and at least one non-prompt lepton with $p_T>10$ GeV. We require at least two light jets with $p_T > 25$ GeV. For all the reconstructed objects (leptons and jets), we require $|\eta_{\ell,j}| < 2.5$.
        
        \item \textbf{Lepton displacement cut:} For the non-prompt leptons we require $|d_0|>2$ mm and $|d_z|<100$ mm. For the prompt leptons we require $|d_0|<0.1$ mm and $|d_z|<1$ mm.

        \item \textbf{Jet pairings:} All possible pairs of jets are considered and for each pair, the dijet invariant mass $m_{jj}$ is computed. Finally, for each dijet pair, the angular separation $\Delta R(\ell^{\text{disp}},jj)$ is calculated with respect to each displaced lepton $\ell^{\text{disp}}$. The dijet candidates are ordered according to increasing $\Delta R(\ell^{\text{disp}},jj)$.

        \item \textbf{$W$ boson reconstruction criteria:} A dijet pair is accepted if it satisfies $m_W - 25~\text{GeV} \leq m_{jj} \leq m_W + 25~\text{GeV}$ and $\Delta R(\ell^{\text{disp}},jj) \leq 2$. The first selection criterion ensures that the jet pair is from the $W$ boson and the second criterion ensures that the reconstructed $W$ boson and the displaced lepton are close enough and therefore coming from the long-lived HNL state.   
        
        \item \textbf{Sequential matching:}  The above selection (item (d)) is first applied to the displaced lepton with the largest $|d_0|$. If no candidate satisfies the criteria, the procedure is repeated for the second displaced lepton if available.
    \end{enumerate}
    \item \textbf{2DV selection criteria:}
\begin{enumerate}

\item \textbf{Preselection:} We require at least three leptons with $p_T >10$ GeV and at least four light jets with $p_T>25$ GeV. For all reconstructed objects (leptons and jets) we require $|\eta_{\ell,j}|<2.5$. 

\item \textbf{Lepton displacement cut:} We require two non-prompt leptons with $|d_0|>2$ mm and $|d_z|<100$ mm. Additionally, we require at least one prompt lepton with $|d_0|<0.1$ mm and $|d_z|<1$ mm.

\item \textbf{Jet pairing:} Out of four leading jets, all possible combinations of two dijet systems are considered. For each combination, the invariant masses of the two dijet systems are calculated.

\item \textbf{$W$ boson reconstruction criteria:} A combination is accepted if both dijet systems satisfy
$m_W-25~\text{GeV} \leq m_{jj} \leq m_W+25~\text{GeV}$.
This requirement ensures that both dijet systems are from $W$ bosons.

\item \textbf{Displaced lepton $W$ matching:} For each dijet system the angular separation $\Delta R(\ell^{\text{disp}},jj)$ is computed with respect to the displaced leptons. The event is accepted if each displaced lepton is associated with one of the reconstructed $W$ bosons satisfying
$\Delta R(\ell^{\text{disp}},jj) \leq 2$ and therefore ensuring that the displaced lepton and the $W$ boson are from the long-lived HNL state.
\end{enumerate}
\end{enumerate}

The SM backgrounds consist of decays of metastable particles such as bottom, charm, and strange hadrons. Other backgrounds can arise in processes such as QCD dijets and $W+\text{jets}$. The latter are processes with large cross sections combined with a very low probability to produce a DV satisfying the selection criteria. Another background arises from displaced photon conversions in the detector material, primarily in the electron mode. The background can also include random crossing tracks from pileup interactions and underlying activity such as multi-parton interactions making fake DVs, nuclear interactions with the detector material, cosmic rays and beam-halo muons. A loose maximal requirement on $|d_z|$ is used for the non-prompt leptons to reduce backgrounds from cosmic ray muons and pileup. Since the background estimates for DV signatures based on Monte Carlo simulations are very challenging, a full simulation of the background is beyond the scope of this work. We presume that the background can be significantly reduced by the selection criteria discussed above\footnote{In experiments looking for DV signatures, the background can be significantly reduced by reconstructing the DV with at least two charged tracks satisfying $\Delta R > 0.1$, a minimum distance of $l_0 > 5$ mm and DV invariant mass $m_{DV}\geq 5$ GeV \cite{Drewes:2019fou}. \textcolor{black}{In fact, events passing the 1DV and 2DV selection criteria predominantly satisfy the requirement $m_{DV}\geq 5$ GeV.}}. 

\section{Results and Discussions}
\label{Sec: Results}

\begin{figure}[t!]
	\includegraphics[width=8.1cm]{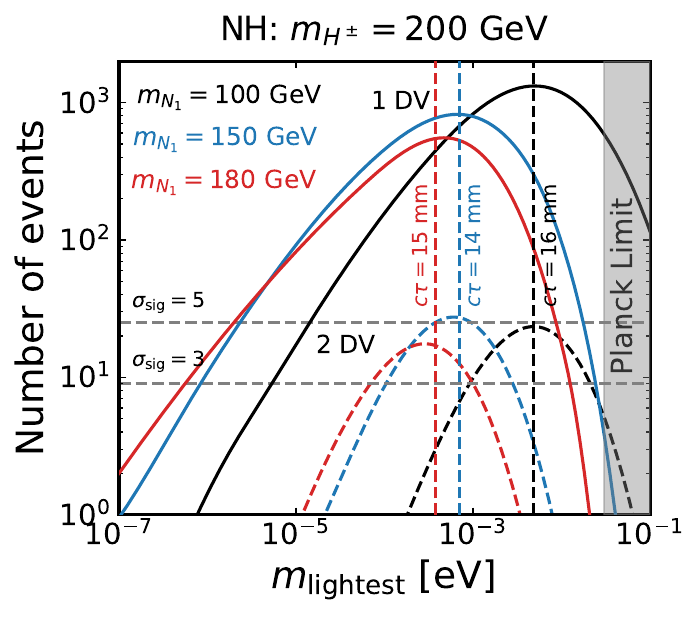}
    \includegraphics[width=8.1cm]{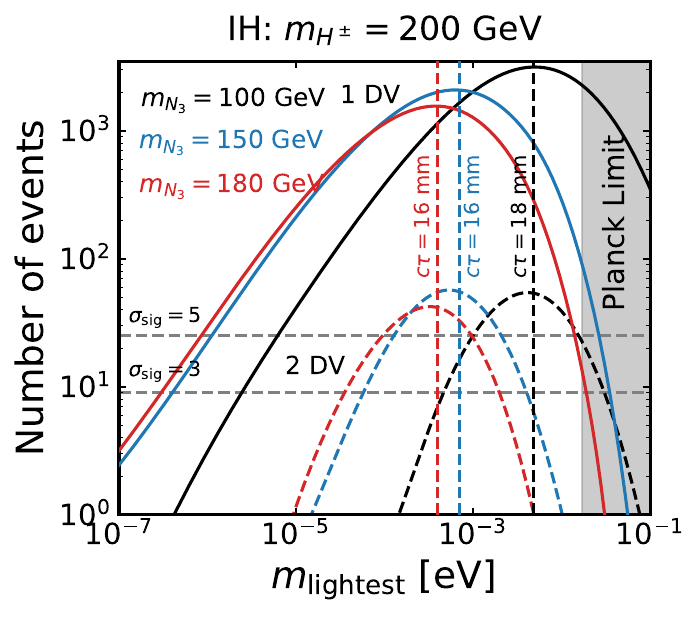}
    \includegraphics[width=8.1cm]{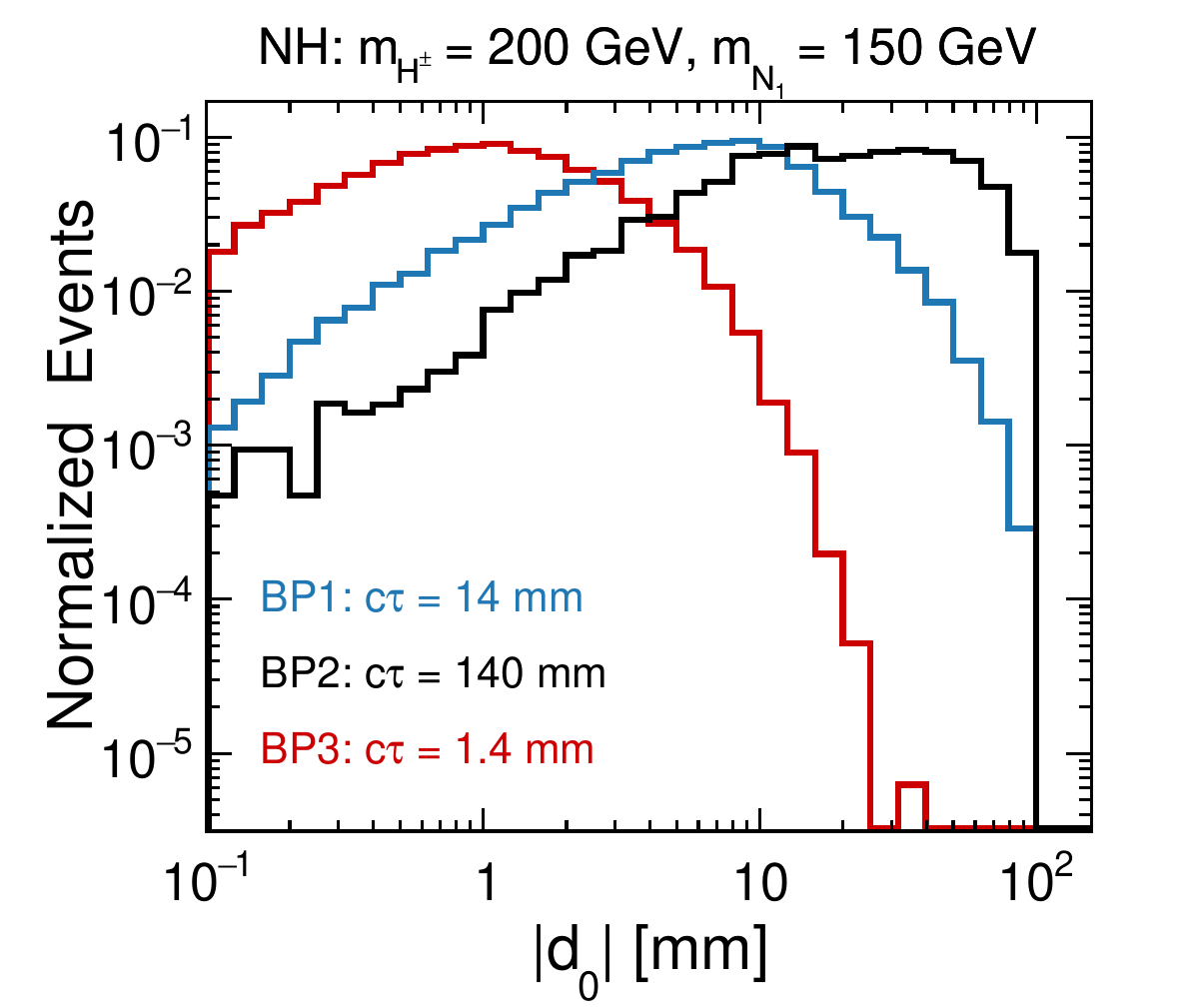}
    \includegraphics[width=8.1cm]{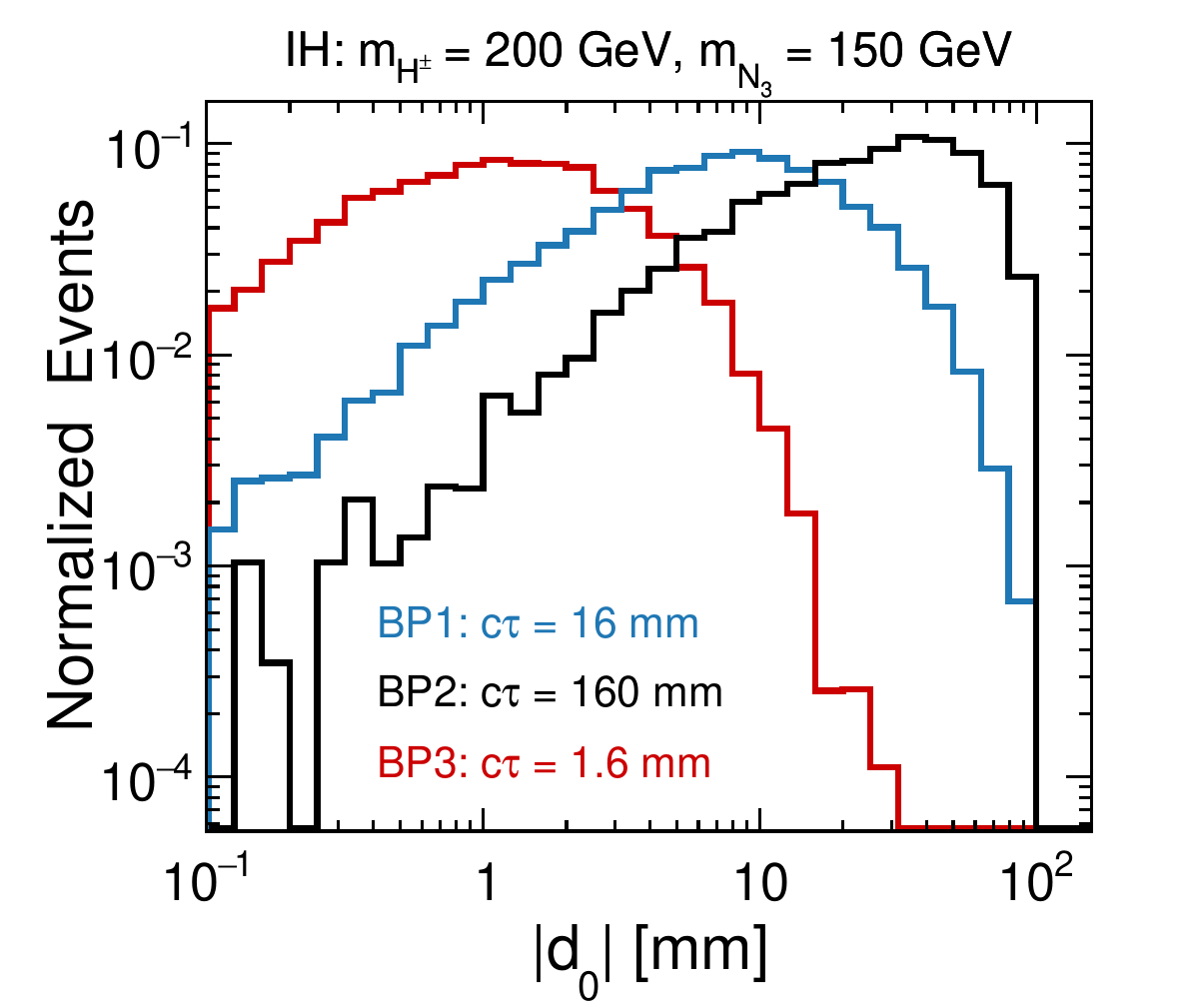}
	\caption{Top: Signal events for the 1DV and 2DV signatures in the NH (left) and IH (right) for $m_{H^\pm}=200$~GeV and long-lived HNL masses $m_{N_{1,3}}=100$, $150$, and $180$~GeV at the 14 TeV LHC with 3000 fb$^{-1}$ luminosity. The vertical shaded region is excluded by the Planck upper limit on the sum of neutrino masses. Bottom: Normalized $|d_0|$ distributions for three benchmark points in the NH (left) and IH (right). BP1 with optimized $c\tau$, BP2 with $c\tau$ ten times larger than in BP1, and BP3 with $c\tau$ one-tenth of that in BP1.}
	\label{Fig: signal events}
\end{figure}

Based on the selection criteria described in Sec.~\ref{Sec: Collider Simulation and Analysis}, we obtain the expected number of signal events for the 1DV and 2DV signatures at the 14 TeV LHC with a luminosity of 3000 fb$^{-1}$  in both the NH and IH scenarios. The top panel of Fig.~\ref{Fig: signal events} shows the number of signal events for the NH case (left) and IH case (right) in the 1DV and 2DV searches. In these analyses, we consider $m_{H^\pm} = 200$~GeV, for which the $H^\pm$ pair production cross section at the 14 TeV LHC is $18$ fb, and long-lived HNL masses $m_{N_{1,3}} = 100$, $150$, and $180$~GeV. We varied the lightest neutrino mass $m_{\text{lightest}}$ in the range $10^{-7}$ eV to $10^{-1}$ eV. Since in 1DV signature we require at least one displaced lepton, the number of signal events surviving the selection criteria is significantly larger than the 2DV signature. The signal events in IH is higher than the signal events in NH.  We also observe that the peaks of the curves corresponding to different choices of the long-lived HNL masses occur at approximately $c\tau \simeq 10$--$20$~mm. The peaks shift toward smaller values of $m_{\text{lightest}}$ as the mass of the long-lived HNL increases. This behavior can be understood from the decay widths of long-lived HNL states, given in Eqs.~[\ref{Eq: HNL decay widths}] and [\ref{Eq: RdR}], and as illustrated in Fig.~\ref{Fig: HNL decay length}. 

The peak positions, as shown in the top panel of Fig.~\ref{Fig: signal events}, arise from the finite size of the tracking volume, the $|d_0|$ efficiency that vanishes at $|d_0|=100$~mm, and the requirement of the lepton displacement cut $|d_0| > 2$~mm. \textcolor{black}{To demonstrate this feature, we consider a benchmark point (BP1) with $m_{H^\pm}=200$~GeV and $m_{N_{1,3}}=150$~GeV. The corresponding proper decay lengths at the peak are $c\tau\simeq14$~mm in the NH case and $c\tau\simeq16$~mm in the IH case. The associated values of the lightest neutrino mass are $m_{\text{lightest}}\simeq7.0\times10^{-4}$~eV for NH and IH.} We consider two additional benchmark points, BP2 for which $c\tau$ is ten times larger than in BP1 and BP3 for which $c\tau$ is one-tenth of that in BP1. The normalized $|d_0|$ distributions are shown in the lower panel in Fig.~\ref{Fig: signal events} with NH in the left and IH in the right. For BP1 (blue), the $|d_0|$ distribution shows a significant number of events beyond 2 mm while remaining largely within the tracking volume, where the $|d_0|$ efficiency is high. For BP2 (black), the larger decay length results in many events with large displacements where the $|d_0|$ efficiency is poor or the decay occurs outside the tracking volume. Finally, for BP3 (red), the smaller decay length leads to most events having small displacements, so the $|d_0|>2$~mm selection cut significantly reduces the number of signal events.  

\begin{figure}[t!]
	\includegraphics[width=8.cm]{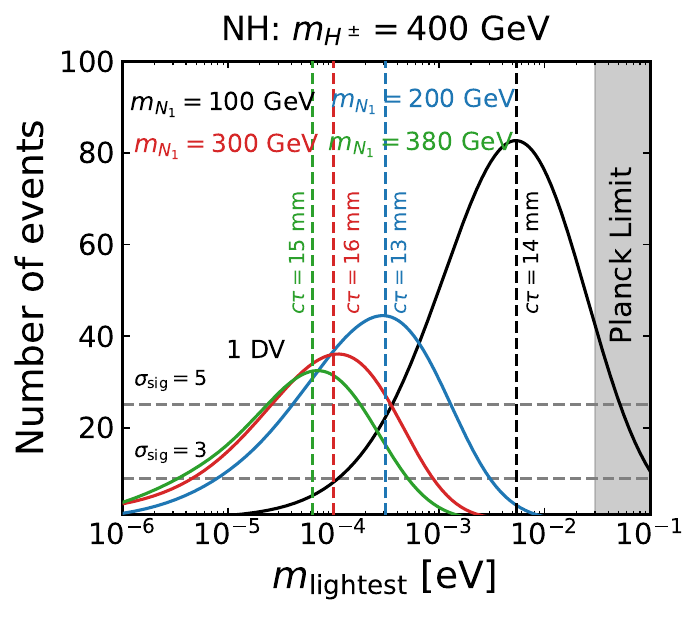}
    \includegraphics[width=8.cm]{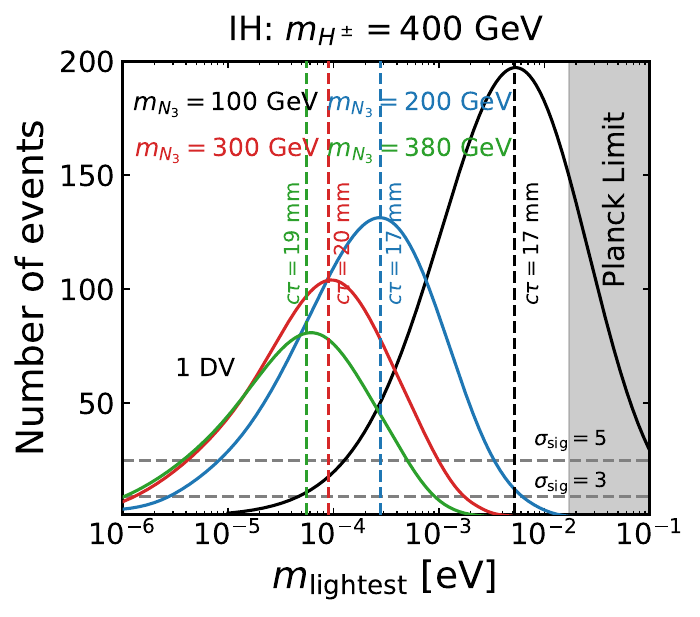}
	\caption{Signal events for the 1DV signature in the NH (left) and IH (right) for $m_{H^\pm}=400$~GeV and long-lived HNL masses $m_{N_{1,3}}=100$, $200$, $300$, and $380$~GeV at the 14 TeV LHC with 3000 fb$^{-1}$ luminosity. The vertical shaded region is excluded by the Planck upper limit on the sum of neutrino masses.}
	\label{Fig: signal events 2}
\end{figure}

We next consider $m_{H^\pm}=400$~GeV, for which the $H^\pm$ pair production cross section at the 14 TeV LHC is approximately $1$~fb. The long-lived HNL masses considered are $m_{N_{1,3}} = 100$, $200$, $300$, and $380$~GeV. Fig.~\ref{Fig: signal events 2} shows the expected number of signal events for the 1DV signature for NH in the left and IH in the right. Once again, the signal peaks occur for $c\tau$ in the range $10$--$20$~mm. In this scenario, the $H^\pm$ pair-production cross section is reduced by a factor of 18 compared to the previous case. Consequently, the total signal yield is also reduced by approximately the same factor. This can be seen by comparing the number of signal events for $m_{N_{1,3}}=100$~GeV in the top panel of Fig.~\ref{Fig: signal events} with those in Fig.~\ref{Fig: signal events 2}. As the $H^\pm$ pair creation cross section is reduced, the sensitivity of the 2DV signature is significantly diminished, and therefore not shown in Fig.~\ref{Fig: signal events 2}.

Since we do not perform a detailed background analysis and assume that the backgrounds are strongly suppressed by the DV selection criteria, we estimate the minimum number of signal events required for a $3\sigma$ evidence and a $5\sigma$ discovery. These are indicated by black dashed horizontal lines in the top panel of Fig.~\ref{Fig: signal events} and in Fig.~\ref{Fig: signal events 2}. We observe that the IH scenario provides better prospects for DV signatures. This difference arises because the long-lived HNL $N_3$ in the IH scenario decays predominantly into muons, and displaced muons have a larger $|d_0|$ dependent efficiency, as given in Eqs.~[\ref{Eq: d0 efficiencies_1}] and [\ref{Eq: d0 efficiencies_2}]. Before we conclude, we briefly discuss the prospects of the 1DV signal events with muon and electron vetoes with $p_T>$10~GeV, which could help to distinguish between the NH and IH neutrino mass hierarchies. The signal events in NH with muon veto and similarly the electron veto in IH will depend on the branching ratios of $H^\pm$ and the long-lived HNL states given in Fig.~\ref{Fig: BR plots}, as well as on the $|d_0|$ efficiencies, reconstruction and identification efficiencies. \textcolor{black}{In the NH a muon veto reduces the signal yield by $40$--$60\%$, while in the IH an electron veto reduces the signal yield by $5$--$15\%$.} From these naive lepton veto estimates, we see that the discrimination appears significantly stronger for IH, because the electron veto in IH reduces the signal event much less than the muon veto does in NH. The 2DV analysis could also be extended to the $4\ell\,4j$ final state with three same sign leptons, indicating LNV by two units and the Majorana nature of the neutrino. However, the already small 2DV signal events would be further suppressed, resulting in a loss of sensitivity.

\section{Conclusions}
\label{Sec: Conclusions}

We have investigated the prospects of long-lived HNL production from charged Higgs decay in a neutrinophilic Higgs doublet model. The small VEV of the neutrinophilic doublet naturally makes the Dirac mass term small, with a sizable Dirac Yukawa coupling, even for a relatively lighter HNL mass scale. Because of this structure the charged Higgs decays predominantly into the HNL states, if kinematically allowed. We considered the long-lived HNL states to be lighter than the charged Higgs, whereas the other two HNLs were considered heavier than the charged Higgs, for simplicity. The long-lived nature in the standard seesaw can be achieved by the appropriate choice of the lightest neutrino in both the hierarchies. However, this long-lived nature cannot be achieved in the minimal seesaw with only two RHNs. 

In our analysis, we considerd charged Higgs pair production at the HL-LHC with $\sqrt{s}=14$ TeV and at an integrated luminosity of 3000 fb$^{-1}$. The charged Higgs decays promptly into the long-lived HNL states and charged leptons, followed by the HNL decay to charged lepton and on-shell $W$ boson, leading to a final state of four leptons and four jets. We potentially can have two displaced vertices, associated with the production of two long-lived HNL states. For our analysis we considered one displaced vertex as well as two displaced vertices signatures. We performed a dedicated simulation to identify the displaced leptons by using the $|d_0|$ dependent efficiencies provided by the CMS collaboration. \textcolor{black}{We showed that for the observation of one displaced vertex, which requires at least one displaced lepton, high statistical significance can be achieved for charged Higgs pair production cross section $>\mathcal{O}(1)$ fb. On the other hand, observing two displaced vertices is more challenging, and the statistical significance decreases substantially for charged Higgs production cross sections below approximately $\mathcal{O}(10)$ fb.}

Finally, we comment on the possibility of enhancing the charged Higgs production cross section. We can consider multi-TeV future lepton colliders like Compact Linear Collider (CLIC) \cite{Adli:2025swq} and muon colliders \cite{InternationalMuonCollider:2025sys} which can produce charged Higgs pair at a cross section of $\mathcal{O}(10)$ fb even for 1~TeV charged Higgs mass.  \textcolor{black}{Neutrinophilic Higgs bosons can also arise in $U(1)$ gauge extension models, where an additional Higgs doublet and the RHNs are charged under the new gauge symmetry. In such models, the extra Higgs doublet generates tiny neutrino Dirac masses through its small VEV. Consequently the charged Higgs originating from the second Higgs doublet is again neutrinophilic. Interestingly, in this setup, charged Higgs pair production at the LHC, mediated by a resonant $Z'$, can remain sizable even for a heavy charged Higgs with mass around the TeV scale \cite{Das:2022cmv}.} Once we achieve a larger cross section, we can study the two displaced vertices signature and possibly even the lepton number violation signal via same signed displaced leptons. 

\section*{Acknowledgments}
\noindent
P.S. thanks KIAS Center for Advanced Computation for providing computing resources.
The work of N.O. is supported in part by the United States Department of Energy Grant
Nos. DE-SC0012447 and DE-SC0026347.
The work of P.S. is supported by Basic Science Research Program through the National Research Foundation of Korea (NRF) funded by the Ministry of Education through the Center for Quantum Spacetime (CQUeST) of Sogang University with grant number RS-2020-NR049598 and by the Ministry of Science and ICT with grant number RS-2025-24523022.

\bibliographystyle{JHEP}
\bibliography{HpHm_HNL}

\end{document}